%Paper: hep-th/9305169
%From: Kanehisa Takasaki <TAKASAKI%JPNYITP.BITNET@pucc.Princeton.EDU>
%Date: Sat, 29 May 93 15:47:31 JST

%%% Nonabelian KP hierarchy with Moyal algebraic coefficients
%%% by Kanehisa Takasaki, Kyoto University KUCP-0062/93
%%%%%%%%%%%%%%%%%%%%%%%%%%%%%%%%%%%%%%%%%%%%%%%%%%%%%%%%%%%%%%
\input phyzzx.tex
%%%%%%%%%%% redefinition of "phyzzx.tex" macros %%%%%%%%%%%%%%%
\catcode`\@=11
\def\wlog#1{}
\def\eqname#1{\rel@x {\pr@tect
  \ifnum\equanumber<0 \xdef#1{{\rm\number-\equanumber}}%
     \gl@bal\advance\equanumber by -1
  \else \gl@bal\advance\equanumber by 1
     \ifx\chapterlabel\rel@x \def\d@t{}\else \def\d@t{.}\fi
    \xdef#1{{\rm\chapterlabel\d@t\number\equanumber}}\fi #1}}
%% "(  )" are removed
%% use vanilla.sty macro "\tag" instead, which can work with "\align"
%% equation name should be cited with "(  )" in the text
%%
\catcode`\@=12
%%%%%%%%%%% macros extracted from "vanilla.sty" %%%%%%%%%%%%%%%
\catcode`\@=11
      %change to CM fonts 3-31-87
%\font\tensmc=amcsc10

%%
%%
\def\eat@#1{}
\mathchardef\prime@="0230
\def\prime{{{}\prime@{}}}
\def\prim@s{\prime@\futurelet\next\pr@m@s}

\def\,{\relax\ifmmode\mskip\thinmuskip\else\thinspace\fi}
\def\!{\relax\ifmmode\mskip-\thinmuskip\else\negthinspace\fi}
\def\frac#1#2{{#1\over#2}}
\def\dfrac#1#2{{\displaystyle{#1\over#2}}}

\def\binom#1#2{{#1\choose #2}}

\def\:{\nobreak\hskip.1111em{:}\hskip.3333em plus .0555em\relax}
\def\intic@{\mathchoice{\hskip5\p@}{\hskip4\p@}{\hskip4\p@}{\hskip4\p@}}
\def\negintic@
 {\mathchoice{\hskip-5\p@}{\hskip-4\p@}{\hskip-4\p@}{\hskip-4\p@}}
\def\intkern@{\mathchoice{\!\!\!}{\!\!}{\!\!}{\!\!}}
\def\intdots@{\mathchoice{\cdots}{{\cdotp}\mkern1.5mu
    {\cdotp}\mkern1.5mu{\cdotp}}{{\cdotp}\mkern1mu{\cdotp}\mkern1mu
      {\cdotp}}{{\cdotp}\mkern1mu{\cdotp}\mkern1mu{\cdotp}}}
\newcount\intno@
\def\iint{\intno@=\tw@\futurelet\next\ints@}
\def\iiint{\intno@=\thr@@\futurelet\next\ints@}
\def\iiiint{\intno@=4 \futurelet\next\ints@}
\def\idotsint{\intno@=\z@\futurelet\next\ints@}
\def\ints@{\findlimits@\ints@@}
\newif\iflimtoken@
\newif\iflimits@
\def\findlimits@{\limtoken@false\limits@false\ifx\next\limits
 \limtoken@true\limits@true
   \else\ifx\next\nolimits\limtoken@true\limits@false
    \fi\fi}
\def\multintlimits@{\intop\ifnum\intno@=\z@\intdots@
  \else\intkern@\fi
    \ifnum\intno@>\tw@\intop\intkern@\fi
     \ifnum\intno@>\thr@@\intop\intkern@\fi\intop}
\def\multint@{\int\ifnum\intno@=\z@\intdots@\else\intkern@\fi
   \ifnum\intno@>\tw@\int\intkern@\fi
    \ifnum\intno@>\thr@@\int\intkern@\fi\int}
\def\ints@@{\iflimtoken@\def\ints@@@{\iflimits@
   \negintic@\mathop{\intic@\multintlimits@}\limits\else
    \multint@\nolimits\fi\eat@}\else
     \def\ints@@@{\multint@\nolimits}\fi\ints@@@}
\def\Sb{_\bgroup\vspace@
        \baselineskip=\fontdimen10 \scriptfont\tw@
        \advance\baselineskip by \fontdimen12 \scriptfont\tw@
        \lineskip=\thr@@\fontdimen8 \scriptfont\thr@@
        \lineskiplimit=\thr@@\fontdimen8 \scriptfont\thr@@
        \Let@\vbox\bgroup\halign\bgroup \hfil$\scriptstyle
            {##}$\hfil\cr}
\def\endSb{\crcr\egroup\egroup\egroup}
\def\Sp{^\bgroup\vspace@
        \baselineskip=\fontdimen10 \scriptfont\tw@
        \advance\baselineskip by \fontdimen12 \scriptfont\tw@
        \lineskip=\thr@@\fontdimen8 \scriptfont\thr@@
        \lineskiplimit=\thr@@\fontdimen8 \scriptfont\thr@@
        \Let@\vbox\bgroup\halign\bgroup \hfil$\scriptstyle
            {##}$\hfil\cr}
\def\endSp{\crcr\egroup\egroup\egroup}
\def\Let@{\relax\iffalse{\fi\let\\=\cr\iffalse}\fi}
\def\vspace@{\def\vspace##1{\noalign{\vskip##1 }}}
\def\aligned{\,\vcenter\bgroup\vspace@\Let@\openup\jot\m@th\ialign
  \bgroup \strut\hfil$\displaystyle{##}$&$\displaystyle{{}##}$\hfil\crcr}
\def\endaligned{\crcr\egroup\egroup}
\def\matrix{\,\vcenter\bgroup\Let@\vspace@
    \normalbaselines
  \m@th\ialign\bgroup\hfil$##$\hfil&&\quad\hfil$##$\hfil\crcr
    \mathstrut\crcr\noalign{\kern-\baselineskip}}
\def\endmatrix{\crcr\mathstrut\crcr\noalign{\kern-\baselineskip}\egroup
                \egroup\,}
\newtoks\hashtoks@
\hashtoks@={#}
\def\format{\crcr\egroup\iffalse{\fi\ifnum`}=0 \fi\format@}
\def\format@#1\\{\def\preamble@{#1}%
  \def\c{\hfil$\the\hashtoks@$\hfil}%
  \def\r{\hfil$\the\hashtoks@$}%
  \def\l{$\the\hashtoks@$\hfil}%
  \setbox\z@=\hbox{\xdef\Preamble@{\preamble@}}\ifnum`{=0 \fi\iffalse}\fi
   \ialign\bgroup\span\Preamble@\crcr}

\def\cases{\left\{\,\vcenter\bgroup\vspace@
     \normalbaselines\openup\jot\m@th
       \Let@\ialign\bgroup$##$\hfil&\quad$##$\hfil\crcr
      \mathstrut\crcr\noalign{\kern-\baselineskip}}

\newif\iftagsleft@
\tagsleft@true
\def\TagsOnRight{\global\tagsleft@false}
\def\tag#1$${\iftagsleft@\leqno\else\eqno\fi
 \hbox{\def\pagebreak{\global\postdisplaypenalty-\@M}%
 \def\nopagebreak{\global\postdisplaypenalty\@M}\rm(#1\unskip)}%
  $$\postdisplaypenalty\z@\ignorespaces}
\interdisplaylinepenalty=\@M
\def\allowdisplaybreak@{\def\allowdisplaybreak{\noalign{\allowbreak}}}
\def\displaybreak@{\def\displaybreak{\noalign{\break}}}
\def\align#1\endalign{\def\tag{&}\vspace@\allowdisplaybreak@\displaybreak@
  \iftagsleft@\lalign@#1\endalign\else
   \ralign@#1\endalign\fi}
\def\ralign@#1\endalign{\displ@y\Let@\tabskip\centering
   \halign to\displaywidth
     {\hfil$\displaystyle{##}$\tabskip=\z@&$\displaystyle{{}##}$\hfil
       \tabskip=\centering&\llap{\hbox{(\rm##\unskip)}}\tabskip\z@\crcr
             #1\crcr}}
\def\lalign@
 #1\endalign{\displ@y\Let@\tabskip\centering\halign to \displaywidth
   {\hfil$\displaystyle{##}$\tabskip=\z@&$\displaystyle{{}##}$\hfil
   \tabskip=\centering&\kern-\displaywidth
        \rlap{\hbox{(\rm##\unskip)}}\tabskip=\displaywidth\crcr
               #1\crcr}}
\def\overrightarrow{\mathpalette\overrightarrow@}
\def\overrightarrow@#1#2{\vbox{\ialign{$##$\cr
    #1{-}\mkern-6mu\cleaders\hbox{$#1\mkern-2mu{-}\mkern-2mu$}\hfill
     \mkern-6mu{\to}\cr
     \noalign{\kern -1\p@\nointerlineskip}
     \hfil#1#2\hfil\cr}}}
\def\overleftarrow{\mathpalette\overleftarrow@}
\def\overleftarrow@#1#2{\vbox{\ialign{$##$\cr
     #1{\leftarrow}\mkern-6mu\cleaders
      \hbox{$#1\mkern-2mu{-}\mkern-2mu$}\hfill
      \mkern-6mu{-}\cr
     \noalign{\kern -1\p@\nointerlineskip}
     \hfil#1#2\hfil\cr}}}
\def\overleftrightarrow{\mathpalette\overleftrightarrow@}
\def\overleftrightarrow@#1#2{\vbox{\ialign{$##$\cr
     #1{\leftarrow}\mkern-6mu\cleaders
       \hbox{$#1\mkern-2mu{-}\mkern-2mu$}\hfill
       \mkern-6mu{\to}\cr
    \noalign{\kern -1\p@\nointerlineskip}
      \hfil#1#2\hfil\cr}}}
\def\underrightarrow{\mathpalette\underrightarrow@}
\def\underrightarrow@#1#2{\vtop{\ialign{$##$\cr
    \hfil#1#2\hfil\cr
     \noalign{\kern -1\p@\nointerlineskip}
    #1{-}\mkern-6mu\cleaders\hbox{$#1\mkern-2mu{-}\mkern-2mu$}\hfill
     \mkern-6mu{\to}\cr}}}
\def\underleftarrow{\mathpalette\underleftarrow@}
\def\underleftarrow@#1#2{\vtop{\ialign{$##$\cr
     \hfil#1#2\hfil\cr
     \noalign{\kern -1\p@\nointerlineskip}
     #1{\leftarrow}\mkern-6mu\cleaders
      \hbox{$#1\mkern-2mu{-}\mkern-2mu$}\hfill
      \mkern-6mu{-}\cr}}}
\def\underleftrightarrow{\mathpalette\underleftrightarrow@}
\def\underleftrightarrow@#1#2{\vtop{\ialign{$##$\cr
      \hfil#1#2\hfil\cr
    \noalign{\kern -1\p@\nointerlineskip}
     #1{\leftarrow}\mkern-6mu\cleaders
       \hbox{$#1\mkern-2mu{-}\mkern-2mu$}\hfill
       \mkern-6mu{\to}\cr}}}
\def\sqrt#1{\radical"270370 {#1}}
\def\dots{\relax\ifmmode\let\next=\ldots\else\let\next=\tdots@\fi\next}
\def\tdots@{\unskip\ \tdots@@}
\def\tdots@@{\futurelet\next\tdots@@@}
\def\tdots@@@{$\mathinner{\ldotp\ldotp\ldotp}\,
   \ifx\next,$\else
   \ifx\next.\,$\else
   \ifx\next;\,$\else
   \ifx\next:\,$\else
   \ifx\next?\,$\else
   \ifx\next!\,$\else
   $ \fi\fi\fi\fi\fi\fi}
\def\text{\relax\ifmmode\let\next=\text@\else\let\next=\text@@\fi\next}
\def\text@@#1{\hbox{#1}}
\def\text@#1{\mathchoice
 {\hbox{\everymath{\displaystyle}\def\textfonti{\the\textfont1 }%
    \def\textfontii{\the\textfont2 }\textdef@@ T#1}}
 {\hbox{\everymath{\textstyle}\def\textfonti{\the\textfont1 }%
    \def\textfontii{\the\textfont2 }\textdef@@ T#1}}
 {\hbox{\everymath{\scriptstyle}\def\textfonti{\the\scriptfont1 }%
   \def\textfontii{\the\scriptfont2 }\textdef@@ S\rm#1}}
 {\hbox{\everymath{\scriptscriptstyle}%
   \def\textfonti{\the\scriptscriptfont1 }%
   \def\textfontii{\the\scriptscriptfont2 }\textdef@@ s\rm#1}}}
\def\textdef@@#1{\textdef@#1\rm \textdef@#1\bf
   \textdef@#1\sl \textdef@#1\it}

\def\textdef@#1#2{%
 \def\next{\csname\expandafter\eat@\string#2fam\endcsname}%
\if S#1\edef#2{\the\scriptfont\next\relax}%
 \else\if s#1\edef#2{\the\scriptscriptfont\next\relax}%
 \else\edef#2{\the\textfont\next\relax}\fi\fi}
\scriptfont\itfam=\tenit \scriptscriptfont\itfam=\tenit
\scriptfont\slfam=\tensl \scriptscriptfont\slfam=\tensl
\mathcode`\0="0030
\mathcode`\1="0031
\mathcode`\2="0032
\mathcode`\3="0033
\mathcode`\4="0034
\mathcode`\5="0035
\mathcode`\6="0036
\mathcode`\7="0037
\mathcode`\8="0038
\mathcode`\9="0039
\def\Cal{\relax\ifmmode\let\next=\Cal@\else
    \def\next{\errmessage{Use \string\Cal\space only in %
      math mode}}\fi\next}
    \def\Cal@#1{{\fam2 #1}}
\def\bold{\relax\ifmmode\let\next=\bold@\else
    \def\next{\errmessage{Use \string\bold\space only in %
      math mode}}\fi\next}
    \def\bold@#1{{\fam\bffam #1}}
\mathchardef\Gamma="0000
\mathchardef\Delta="0001
\mathchardef\Theta="0002
\mathchardef\Lambda="0003
\mathchardef\Xi="0004
\mathchardef\Pi="0005
\mathchardef\Sigma="0006
\mathchardef\Upsilon="0007
\mathchardef\Phi="0008
\mathchardef\Psi="0009
\mathchardef\Omega="000A
\mathchardef\varGamma="0100
\mathchardef\varDelta="0101
\mathchardef\varTheta="0102
\mathchardef\varLambda="0103
\mathchardef\varXi="0104
\mathchardef\varPi="0105
\mathchardef\varSigma="0106
\mathchardef\varUpsilon="0107
\mathchardef\varPhi="0108
\mathchardef\varPsi="0109
\mathchardef\varOmega="010A
\def\wlog#1{\immediate\write-1{#1}}
%%\catcode`\@=\active
\catcode`\@=12  %% defining '@' as a letter
%%%%%%%%%%%%%%%%%%%%%%%%%%%%%%%%%%%%%%%%%%%%%%%%%%%%%%%%%%%%%%%%%%
\def\=def{\; \mathop{=}_{\text{\rm def}} \;}
\def\rd{\partial}

\def\phat{\hat{p}}
\def\qhat{\hat{q}}

\def\What{\hat{W}}

\def\what{\hat{w}}
\def\calU{{\cal U}}
\def\calV{{\cal V}}

\def\calL{{\cal L}}
\def\calM{{\cal M}}
\def\calG{{\cal G}}
\def\calB{{\cal B}}
\def\calC{{\cal C}}
\def\calD{{\cal D}}

\def\sl{\text{sl}}
\def\gl{\text{gl}}
\def\sdiff{\text{sdiff}}
\def\Poisson{\text{Poisson}}
\def\Moyal{\text{Moyal}}
\def\[{ \{\!\{ }
\def\]{ \}\!\} }
\def\mapright#1{ \smash{\mathop{\longrightarrow}\limits^{#1}} }
\def\mapdown#1{ \Bigl\downarrow%
  \rlap{$\vcenter{\hbox{$\scriptstyle #1$}}$} }
%%%%%%%%%%%%%%%%%%%%%%%%%%%%%%%%%%%%%%%%%%%%%%%%%%%%%%%%%%%%%%%%%%
\hsize=15.5truecm
\vsize=23truecm
\doublespace
\TagsOnRight
\overfullrule=0pt
%%%%%%%%%%%%%%%%%%%%%%%%%%%%%%%%%%%%%%%%%%%%%%%%%%%%%%%%%%%%%%%%%%
\pubnum={KUCP-0062/93}
\date={May, 1993}
\titlepage
\title{\fourteencp
  Nonabelian \ KP hierarchy \ with
  \break Moyal \ algebraic \ coefficients}
\author{Kanehisa Takasaki}
\address{
  Department of Fundamental Sciences\break
  Faculty of Integrated Human Studies, Kyoto University\break
  Yoshida-Nihonmatsu-cho, Sakyo-ku, Kyoto 606, Japan\break
  E-mail: takasaki @ jpnyitp (Bitnet)\break}

\abstract
\noindent
A higher dimensional analogue of the KP hierarchy is presented.
Fundamental constituents of the theory are pseudo-differential
operators with Moyal algebraic coefficients. The new hierarchy can be
interpreted as large-$N$ limit of multi-component ($\gl(N)$ symmetric)
KP hierarchies. Actually, two different hierarchies are constructed.
The first hierarchy consists of commuting flows and may be thought of
as a straightforward extension of the ordinary and multi-component KP
hierarchies. The second one is a hierarchy of noncommuting flows, and
related to Moyal algebraic deformations of selfdual gravity.  Both
hierarchies turn out to possess quasi-classical limit, replacing
Moyal algebraic structures by Poisson algebraic structures. The
language of W-infinity algebras provides a unified point of view to
these results.
\endpage
%%%%%%%%%%%%%%%%%%%%%%%%%%%%%%%%%%%%%%%%%%%%%%%%%%%%%%%%%%%%%%%%
\chapter{Introduction}

\noindent
The technique of Large-$N$ limit is a magic that produces new
continuous dimensions out of finite matrix indices.  According
to observations since the early eighties
[\REF\SUinfty{
  J. Hoppe,
  Quantum theory of a massless relativistic surface,
  Ph.D. thesis (MIT, 1982);
  Elem. Part. Res. J. (Kyoto) 80 (1989), 145-202. \nextline
  Fairlie, D., Fletcher, P., and Zachos, C.N.,
  Trigonometric structure constants for new infinite algebras,
  Phys. Lett. B218 (1989), 203.\nextline
  Fairlie, D.B., and Zachos, C.K.,
  Infinite-dimensional algebras, sine brackets, and SU($\infty$),
  Phys. Lett. 224B (1989), 101-107.\nextline
  Pope, C.N., and Stelle, K.S.,
  SU$(\infty)$, SU$_+(\infty)$ and area-preserving algebras,
  Phys. Lett. 226B (1989), 257-263.\nextline
  Hoppe, J.,
  Diffeomorphism groups, quantization, and SU($\infty$),
  Int. J. Mod. Phys. A 4 (19) (1989), 5235-5248.
}\SUinfty],
large-$N$ limit of $\sl(N)$ looks like the algebra of infinitesimal
area-preserving (or symplectic) diffeomorphisms on a two dimensional
surface $\Sigma$:
$$
  \lim_{N \to \infty} \sl(N) \simeq \sdiff(\Sigma).  \tag\eq
$$
If central elements (scalar matrices) are added, the limit
becomes a Poisson algebra, i.e., the Lie algebra realized by
a Poisson bracket on the surface:
$$
  \lim_{N \to \infty} \gl(N) \simeq \Poisson(\Sigma). \tag\eq
$$
These intriguing observations are based on the existence of
the so called ``sine generators" in these matrix Lie algebras.
The theory of sine generators further tells us [\SUinfty] that
this is just a special, rather singular case; in general, the
limit can also become a Moyal algebra:
$$
  \lim_{N \to \infty} \gl(N) \simeq \Moyal(\Sigma).  \tag\eq
$$
Moyal algebras are a kind of ``quantum deformation" of Poisson
algebras, which is a unique deformation subject to a set of
natural requirements
[\REF\MoyalAlg{
  Bayen, F., Flato, M., Fronstal, C., Lichnerowicz, A.,
  and Sternheimer, D.,
  Deformation theory and quantization.  1.
  Deformations of symplectic structures,
  Ann. Phys. (N.Y.) 111 (1978), 61;
  ditto 2. Physical applications,
  Ann. Phys. (N.Y.) 111 (1978), 111.\nextline
  Arveson, W.,
  Quantization and uniqueness of invariant structures,
  Commun. Math. Phys. Phys. 89 (1983), 77-102.\nextline
  Fletcher, P., The uniqueness of the Moyal algebra,
  Phys. Lett. B248 (1990), 323-328.\nextline
}\MoyalAlg].

These observations, although mathematically slightly problematical
[\REF\MoyalProb{
  Bordemann, M., Hoppe, J., Schaller, P., and Schlichenmaier, M.,
  $gl(\infty)$ and geometric quantization,
  Commun. Math. Phys. 138 (1991), 209-244.
}\MoyalProb],
have been a very useful point of view for understanding higher
dimensional integrable systems. For instance, large-$N$ limit of
the $\sl(N)$ Toda field theory becomes a three dimensional
nonlinear system (the dispersionless Toda equation)
[\REF\ThreeDToda{
  Saveliev, M.V., and Vershik, A.M.,
  Continual analogues of contragredient Lie algebras,
  Commun. Math. Phys. 126 (1989), 367-378;
  New examples of continuum graded Lie algebras,
  Phys. Lett. 143A (1990), 121-128.\nextline
  Bakas, I.,
  The structure of the $W_\infty$ algebra,
  Commun. Math. Phys. 134 (1990), 487-508.\nextline
  Park, Q-H.,
  Extended conformal symmetries in real heavens,
  Phys. Lett. 236B (1990), 429-432.
}\ThreeDToda].
Similarly, large-$N$ limit of two dimensional nonlinear sigma
models with the pure WZW Lagrangian coincides with selfdual gravity
(the selfdual vacuum Einstein equation)
[\REF\PaSDG{
  Park, Q-H.,
  Self-dual gravity as a large-$N$ limit of
  the 2D non-linear sigma model,
  Phys. Lett. 238B (1990), 287-290;
  2-D sigma model approach to 4-D instantons,
  Int. J. Mod. Phys. A7 (1992), 1415-1448.
}\PaSDG].
These two models are related to Poisson algebras, and well known
to be integrable. Recently, a Moyal algebraic analogue of selfdual
gravity is discovered from the same standpoint
[\REF\StMoyalSDG{
  Strachan, I.A.B.,
  The Moyal algebra and integrable deformations
  of the self-dual Einstein equations,
  Phys. Lett. B282 (1992), 63-66.
}\StMoyalSDG]
and shown to be integrable
[\REF\TaMoyalSDG{
  Takasaki, K.,
  Dressing operator approach to Moyal algebraic
  deformation   of selfdual gravity,
  Kyoto University KUCP-0054/92 (December, 1992).
}\TaMoyalSDG].

Inspired by this point of view, we present in this paper a higher
dimensional analogue of the KP hierarchy.  The celebrated KP
hierarchy is known to possess a $\gl(N)$ symmetric version, the
so called ``$N$-component KP hierarchy"
[\REF\KPGen{
  Sato, M., and Sato, Y.,
  Soliton equations as dynamical systems in
  an infinite dimensional Grassmann manifold,
  in: P.D. Lax et al. (eds.),
  {\it Nonlinear Partial Differential Equations in Applied Sciences}
  (North-Holland, Amsterdam, 1982).\nextline
  Date, E., Jimbo, M., Kashiwara, M., and Miwa, T.,
  Transformation groups for soliton equations,
  in: M. Jimbo and T. Miwa (eds.),
  {\it Nonlinear Integrable Systems ---
  Classical Theory and Quantum Theory}
  (World Scientific, Singapore, 1983).
}\KPGen].
Our new hierarchy may be thought of as a kind of large-$N$ limit
of the $N$-component KP hierarchy. Actually, rather than literally
considering such limit, we directly construct the hierarchy
replacing $\gl(N)$ by the Moyal algebra $\Moyal(M)$ on a symplectic
manifold $M$. For simplicity, we mostly deal with the planar case
(i.e., the case where $M$ is a two dimensional symplectic manifold),
but all results can be extended straightforward to more general cases.

The new hierarchy (which we call the ``nonabelian KP hierarchy with
Moyal algebraic coefficients") turns out to possess several novel
characteristics. Of particular interest is the fact that there are two
distinct types of hierarchies within this framework. The first type of
hierarchy consists of a commuting set of flows like many other
integrable hierarchies.  This hierarchy may be thought of as naive
large-$N$ limit of the $N$-component KP hierarchy. The second type,
meanwhile, is a hierarchy of noncommuting flows.  This noncommuting
hierarchy has a reduction to a hierarchy of integrable flows including
Moyal algebraic deformations of selfdual gravity.  These Moyal
algebraic deformations are somewhat distinct from Strachan's
deformation.

Another large-$N$ limit, i.e., Poisson algebraic limit, can be
realizes as quasi-classical ($\hbar \to 0$) limit of these Moyal
algebraic hierarchies.  Quasi-classical limit of the commuting
hierarchy gives a higher dimensional extension of the dispersionless
KP hierarchy
[\REF\dKPGen{
  Lebedev, D., and Manin, Yu.,
  Conservation laws and Lax representation on
  Benny's long wave equations,
  Phys. Lett. 74A (1979), 154-156.\nextline
  Kodama, Y.,
  A method for solving the dispersionless KP equation
  and its exact solutions,
  Phys. Lett. 129A (1988), 223-226.\nextline
  Kodama, Y., and Gibbons, J.,
  A method for solving the dispersionless KP hierarchy
  and its exact solutions, II,
  Phys. Lett. 135A (1989), 167-170.
}\dKPGen].
Similarly, quasi-classical limit of the noncommuting hierarchy
is related to hierarchies of integrable flows in ordinary selfdual
gravity
[\REF\TaSDG{
  Takasaki, K.,
  An infinite number of hidden variables in hyper-K\"{a}hler metrics,
  J. Math. Phys. 30 (1989), 1515-1521;
  Symmetries of hyper-K\"ahler (or Poisson gauge field) hierarchy,
  ibid 31 (1990), 1877-1888.
}\TaSDG].

This paper is organized as follows.  Sections 2 and 3 are intended
to be a review of key ideas and tools in the theory of integrable
hierarchies and Moyal algebras. The contents of these sections, in
particular Section 2, are mostly model-independent.  Sections 4 and 5
are specialized to the Moyal algebraic hierarchies. Relation to Moyal
algebraic deformations of selfdual gravity is discussed in Section 6.
Quasi-classical limit is treated in Section 7.  Section 8 is a summary
of results from the point of view of W-infinity algebras.

\chapter{General structure of integrable hierarchies}

\noindent
We here present a general framework for dealing with various
integrable hierarchies on an equal footing.  A prototype is the
standard treatment of the ordinary and multi-component KP
hierarchies [\KPGen], which is now reformulated in an abstract
and model-independent way.

\section{Lax and zero-curvature representations}

Integrable hierarchies usually possesses two distinct
representations --- the Lax representation and the zero-curvature
(or Zakharov-Shabat) representation.  Algebraic structures of
these representations are governed by a Lie algebra
(of operators or of matrices) with a direct sum
decomposition into two subalgebras:
$$
  \calG = \calG_{+} \oplus \calG_{-}.
                                       \tag\eqname\DirectSum
$$
Let $(\quad)_{\pm}$ denote the projection onto each component.

Lax and zero-curvature representations consist of differential
equations (``Lax equations" and ``zero-curvature equations") for
$t$-dependent elements $L_\sigma$ and $B_i$ of the above Lie
algebras:
$$
  L_\sigma = L_\sigma(t) \in \calG, \quad
  B_i = B_i(t) \in \calG_{+},
                                                      \tag\eq
$$
where $\sigma$ ranges over a finite index set, $i$ over another
(finite or infinite) index set. Let us call $L_\sigma$'s ``Lax
operators" and $B_i$'s ``Zakharov-Shabat operators."  The Lax
repretation can be written
$$
  \dfrac{\rd L_\sigma}{\rd t_i} = [ B_i,L_\sigma ],
                                                      \tag\eq
$$ where $t_i$ denotes a time variable of the ``$i$-th" flow.
The zero-curvature equations (or ``Zakharov-Shabat equations")
can similarly be written
$$
  \dfrac{\rd B_i}{\rd t_j}
  - \dfrac{\rd B_j}{\rd t_i} + [ B_i, B_j ] = 0.
                                                      \tag\eq
$$
The Lax and Zakharov-Shabat operators should be selected so that
these equations be consistent.

Example (KP hierarchy): For the KP hierarchy, $\calG$ is comprised
of pseudo-differential operators (of finite order) in one variable
$x$ with scalar coefficients,
$$
\align
  & \calG = \{ A \mid A = \sum a_n \rd_x^n \}, \quad
    \rd_x = \rd/\rd x,                                   \\
  & \left( \sum a_n \rd_x^n \right)_{+}
    = \left( \sum a_n \rd_x^n\right)_{\ge 0}
    = \sum_{n \ge 0} a_n \rd_x^n,                        \\
  & \left( \sum a_n\rd_x^n \right)_{-}
    = \left( \sum a_n \rd_x^n \right)_{\le -1}
    = \sum_{n \le -1} a_n \rd_x^n,
                                                 \tag\eq \\
\endalign
$$
The Lax and zero-curvature representations are formulated in
terms of a single Lax operator $L$ of the form
$$
    L = \rd_x + \sum_{n=1}^\infty g_{n+1} \rd_x^{-n} \tag\eq
$$
and and an infinite number of Zakharov-Shabat operators
$B_n$, $n=1,2,\ldots$ given by
$$
  B_n = \left( L^n \right)_{\ge 0}.  \tag\eq
$$

Example (multi-component KP hierarchy): The formulation of the
$N$-component KP hierarchy is very similar, but based on
pseudo-differential operators with $N \times N$ matrix-valued
coefficients.  Besides an analogue $L$ of the Lax operator in
the one-component KP hierarchy, we need $N$ additional Lax
operators $U_\alpha$, $\alpha=1,\ldots,N$ to formulate the Lax
and zero-curvature representations. Flows are numbered by a double
index $i=(n,\alpha)$, ($n=1,2,\ldots$, $\alpha=1,\ldots,N$) and
generated by the Zakharov-Shabat operators
$$
  B_{n\alpha} = \left( L^n U_\alpha \right)_{\ge 0}.  \tag\eq
$$

\section{Dressing operator}

Integrable hierarchies of the above type allows another fundamental
representation --- the dressing operator representation. The dressing
operator is a $t$-dependent group element
$$
  W = W(t) \in \exp\; \calG_{-}                        \tag\eq
$$
of the subalgebra $\calG_{-}$ with which the Lax operators
$L_\sigma$ are expressed in the ``dressing" form
$$
  L_\sigma = W C_\sigma W^{-1},                        \tag\eq
$$
where $C_\alpha$ are ``undressed" operators that are independent
of solutions in consideration.  With a suitable choice of the
dressing operator, the Lax and zero-curvature equations can be
converted into a set of differential equations of the form
$$
  \dfrac{\rd W}{\rd t_i} = B_i W - W G_i,
                                           \tag\eqname\EqOfWTI
$$
where $G_i$ are a commuting set of operators that are also
independent of solutions of the hierarchy. The Zakharov-Shabat
operators turn out to be written
$$
  B_i = \left( W G_i W^{-1} \right)_{+},              \tag\eq
$$
so that one can eliminate $B_i$ from the above equations.
This results in the
equations
$$
  \dfrac{\rd W}{\rd t_i} = - \left( W G_i W^{-1}\right)_{-} W.
                                                       \tag\eq
$$
The last equations give a well defined set of flows on the group
$\exp\; \calG_{-}$ generated by $\calG_{-}$.

Example (KP hierarchy): In the case of the one- and multi-component
KP hierarchy, $W$ is a pseudo-differential operator of the form
$$
  W = 1 + \sum_{n=1}^\infty w_n \rd_x^{-n}.  \tag\eq
$$

Remark: In the traditional theory of integrable hierarchies,
$C_\sigma$'s are usually $t$-independent. We shall, however,
see in later sections that this is too restrictive.
Furthermore, we shall extend the present setting to a case
where $G_i$'s are noncommutative.

\section{Factorization relation}

The above flows on $\exp\; \calG_{-}$ can be translated into a kind
of ``factorization relation" in the group $\exp\; \calG$ generated
by $\calG$ (also called a ``Riemann-Hilbert problem"). The following
formulation is inspired by Mulase's work on the KP hierarchy
[\REF\MuKP{
  Mulase, M.,
  Complete integrability of the Kadomtsev-Petviashvili equation,
  Advances in Math. 54 (1984), 57-66.
}\MuKP].
[Remark: In general, such a group $\exp\;\calG$ might not exist
in a mathematically rigorous sense, but some realization as a
``formal group" is always available as Mulase illustrates for the
case of the KP hierarchy. Because of this, we use this somewhat
loose notation and dare to call it a ``group."]

A fundamental factorization relation can be written
$$
  W(t)E(t)W(0)^{-1} = \What(t) \in\exp\;\calG_{+},
                                         \tag\eqname\FactorW
$$
where $W(0)$ is the ``initial value" of $W = W(t)$ at $t=0$,
$\What(t)$ is a $t$-dependent element of $\exp\;\calG_{+}$,
and $E(t)$ is given by
$$
  E(t) = \exp \left( \sum t_i G_i \right).
                                         \tag\eqname\DefOfE
$$
(To be more precise, one has to enlarge $\calG_{+}$ into a ``formal
completion" with respect to time variables [\MuKP].) The above
relation can be rewritten
$$
  E(t) W(0)^{-1} = W(t)^{-1} \What(t),             \tag\eq
$$
which looks more like a ``factorization" or a ``Riemann-Hilbert
problem."  The direct sum decomposition at the level of the Lie
algebra, (\DirectSum), yields the direct product decomposition
$$
  \exp\; \calG \simeq \exp\; \calG_{-} \times \exp\; \calG_{+}
                                                   \tag\eq
$$
at the group level. The factorizability is thus ensured at least
in a neighborhood of $t=0$.  Eventually, the flows on the space
of dressing operators are converted into the action of $E(t)$ on
the coset space
$$
\matrix
    \exp\; \calG_{-} &\simeq  & \exp\; \calG / \exp\; \calG_{+}  \\
          W          &\mapsto & [W] = W \exp\; \calG_{+}         \\
\endmatrix
                                                       \tag\eq
$$

Let us briefly show how to prove the equivalence of (\EqOfWTI)
and (\FactorW). (For more details, see Mulase's paper [\MuKP]).

(\FactorW) $\Longrightarrow$ (\EqOfWTI):
Differentiating both hand sides of (\FactorW) and recalling
(\DefOfE), one can derive the relation
$$
  \left( \dfrac{\rd W(t)}{\rd t_i}
  + W(t) G_i \right) E(t)W(0)^{-1}
  = \dfrac{\rd \What(t) }{\rd t_i}.                   \tag\eq
$$
Using the factorization relation once again, one can eliminate
the initial data $W(0)$ to obtain
$$
  \left( \dfrac{\rd W(t)}{\rd t_i}
  + W(t) G_i \right) W(t)^{-1}
  = \dfrac{\rd \What(t) }{\rd t_i} \What(t)^{-1}.    \tag\eq
$$
Since the right hand side gives an element of $\calG_{+}$,
the $\calG_{-}$ component of the left hand side should be
identically zero. This gives Eq. (\EqOfWTI).

(\EqOfWTI) $\Longrightarrow$ (\FactorW):
Eq. (\EqOfWTI) can be rewritten
$$
  \dfrac{\rd}{\rd t_i} \bigl( W(t)E(t) \bigr)
  = B_i(t)W(t)E(t),                                  \tag\eq
$$
and repeated use of this formula yields a similar expression
for higher derivatives,
$$
  \dfrac{\rd}{\rd t_{i_1}}
    \cdots \dfrac{\rd}{\rd t_{i_k}}
      \bigl( W(t)E(t) \bigr)
  = B_{i_1 \cdots i_k}(t)W(t)E(t),                   \tag\eq
$$
where $B_{i_1 \cdots i_k}$ lies in $\calG_{+}$.  Evaluated at
$t = 0$, they give Taylor coefficients of $W(t)E(t)$,
$$
\align
  W(t)E(t) =& W(0) + \sum_{i} t_i B_i (0) W(0) + \cdots     \\
           =& \left( 1 + \sum_{i} t_i B_i(0)
                                  + \cdots \right) W(0),
                                                    \tag\eq \\
\endalign
$$
and the first factor $1 + \sum t_i B_i(0) + \cdots$ on the
right hand side becomes an element of $\exp \calG_{+}$.
Writing this factor $\What(t)$, one finds that the last
relation is nothing but the factorization relation.

\chapter{Pseudo-differential operators with
Moyal algebraic coefficients}

\noindent
Basic constituents of our new hierarchies, too, are
pseudo-differential operators.  Pseudo-differential operators
in the $N$-component KP hierarchy have $N \times N$ matrix-valued
coefficients.  We replace these coefficients by elements of a Moyal
algebra $\Moyal(M)$.  For simplicity, we here deal with the simplest
case where $M$ is a two dimensional plane.

\section{Moyal bracket and star product}

Let $(y,z)$ be a pair of canonical coordinates on a two dimensional
planar symplectic manifold. The Moyal algebra consists of functions
(or of formal power series) of $(y,z)$ equipped with the Moyal bracket
$$
  \{ a, b \}_\hbar
  = \frac{2}{\hbar} \left. \sinh \left[
        \frac{\hbar}{2} \left(
           \frac{\rd^2}{\rd y \rd z'}
          -\frac{\rd^2}{\rd z \rd y'} \right)
    \right] a(y,z) b(y',z') \right|_{y'=y,z'=z},
                                   \tag\eqname\MoyalBracket
$$
where $\hbar$ is a nonzero parameter (``Planck constant").
Actually, we rather wish to consider the Moyal algebra of formal
power series of $(y,z)$ (and of $\hbar$). This is indeed possible,
because the right hand side of the above formula is still meaningful
in such a case. The ``Planck constant" $\hbar$ then plays only a
formal role. [Remark: The above definition of the Moyal bracket is
somewhat unusual; usually, ``$\sinh$" is replaced by ``$\sin$".
This is just for simplifying notations, and one can go back to
the ordinary situation by replacing $\hbar \to i \hbar$.]

The Moyal bracket is a kind of ``quantum deformation" of the
Poisson bracket
$$
  \{ a, b \}
  = \frac{\rd a}{\rd y} \frac{\rd b}{\rd z}
   -\frac{\rd a}{\rd z} \frac{\rd b}{\rd y},
                                   \tag\eqname\PoissonBracket
$$
reducing to the latter in the classical limit,
$$
  \{ a, b \}_\hbar \longrightarrow \{ a, b \} \quad
  (\hbar \to 0).                                    \tag\eq
$$
The canonical coordinates $(y,z)$ are indeed canonical for both
these Lie brackets:
$$
  \{ y, z \}_\hbar = 1, \quad \{ y, z \} = 1.       \tag\eq
$$
Unlike the Poisson bracket, however, the Moyal bracket can be written
as a (normalized) commutator
$$
  \{ a, b \}_\hbar = \hbar^{-1} ( a*b - b*a )       \tag\eq
$$
of an associative product --- the star product [\SUinfty][\MoyalAlg]
$$
   a*b = \left. \exp \left[
          \frac{\hbar}{2}\left(
            \frac{\rd^2}{\rd y \rd z'}
           -\frac{\rd^2}{\rd z \rd y'} \right)
          \right] a(y,z) b(y',z') \right|_{y'=y,z'=z}.
                                                    \tag\eq
$$
The star product is nothing but the composition rule of Weyl-ordered
operators.

\section{Pseudo-differential operators with Moyal algebraic
coefficients}

Pseudo-differential operators with coefficients in the Moyal
(or star product) algebra are, by definition, linear combinations
$$
  A = \sum_{n=-\infty}^m a_n \rd_x^n              \tag\eq
$$
of powers of $\rd_x$ with $x$-dependent coefficients taken from
the Moyal algebra:
$$
  a_n = a_n(\hbar,x,y,z).                         \tag\eq
$$
Multiplication of two pseudo-differential operators, which we
write $A*B$ and consider an extension of star product, is defined
term-by-term as:
$$
  ( a \rd_x^m ) * ( b\rd_x^n)
  = \sum_{k=0}^\infty \binom{n}{k}
     a * \frac{\rd^k b}{\rd x^k} \rd_x^{m+n-k}.
                                                  \tag\eq
$$
Actually, it is more natural to expand pseudo-differential
operators in powers of $\hbar \rd_x$ rather than $\rd_x$.
This indeed becomes crucial in the construction of integrable
hierarchies, as we now show.

We now specify a Lie algebra $\calG$ with direct sum decomposition
into two subalgebras for our new hierarchies.  The Lie algebra $\calG$,
by definition, consists of pseudo-differential operators of the form
$$
  A = \hbar^{-1} \sum_{n=-\infty}^m a_n(\hbar\rd_x)^n
                                                 \tag\eq
$$
with Moyal algebraic coefficients that behave smoothly as
$\hbar \to 0$:
$$
  a_n = a_n^{(0)}(x,y,z) + O(\hbar) \quad
  (\hbar \to 0).                                 \tag\eq
$$
It is not hard to check that these pseudo-differential operators
are indeed closed under star product commutator $[A,B] = A*B - B*A$.
The direct sum decomposition is basically the same as in the
multi-component KP hierarchies:
$$
\align
         \calG =& \calG_{\ge 0} \oplus \calG_{\le -1},         \\
   (\quad)_{+} =& (\quad)_{\ge 0}: \ \text{projection onto}\
                  (\hbar\rd_x)^0, (\hbar\rd_x)^1,\ldots,
                                                               \\
   (\quad)_{-} =& (\quad)_{\le -1}: \ \text{projection onto}\
                  (\hbar\rd_x)^{-1}, (\hbar\rd_x)^{-2},\ldots.
                                                       \tag\eq \\
\endalign
$$
The dressing operator should be a group element of $\calG_{\le -1}$:
$$
  W = \exp_{*} \hbar^{-1} A, \quad
  A \in \calG_{\le -1}.
                                                     \tag\eq
$$
One can further expand $W$ into negative powers of $\hbar\rd_x$,
$$
  W = 1 + \sum_{n=1}^\infty w_n (\hbar\rd_x)^{-n},
                                                     \tag\eq
$$
but the coefficients $w_n$ will have singularities at $\hbar = 0$
that are rather hard to control.  This is by no means a special
property of the present situation, but common to this kind of
formulation of $\hbar$-dependent integrable hierarchies.

\chapter{Hierarchy of commuting flows}

\noindent
We now proceed to the construction of the first hierarchy that
consists of commuting flows.  The construction is almost parallel
to the ordinary and multi-component KP hierarchies.

\section{Equations of dressing operator}

Let us first formulate the hierarchy in the language of the operator
$W$ with Moyal algebraic coefficients. The Lie algebra $\calG$ and
its direct sum decomposition have been specified in the end of the
last section. As in the case of the $N$-component KP hierarchy, the
index $i$ of flows is a double index $(n,\alpha)$. Both components,
however, range over all nonnegative integers as $n = 0,1,2,\ldots$,
$\alpha=0,1,2,\ldots$. Generators $G_i$ of flows are given by
$$
  G_{n\alpha} = (\hbar \rd_x)^n y^\alpha,         \tag\eq
$$
which obviously commute with each other. Differential equations of
the dressing operator are given by
$$
\align
  & \hbar\dfrac{\rd W}{\rd t_{n\alpha}}
     = B_{n\alpha}*W - W*G_{n\alpha},                           \\
  & B_{n\alpha}
     = \left( W * G_{n\alpha} * W^{-1} \right)_{\ge 0}.
                                                        \tag\eq \\
\endalign
$$
To see that these equations fall into the general framework
of the previous sections, let us rewrite these equations as
$$
\align
  & \dfrac{\rd W}{\rd t_{n\alpha}}
     = \hbar^{-1}B_{n\alpha}*W - W*\hbar^{-1}G_{n\alpha},     \\
  & \hbar^{-1}B_{n\alpha}
     = \left( W * \hbar^{-1}G_{n\alpha} * W^{-1} \right)_{\ge 0}.
                                         \tag\eqname\EqOfWTNA \\
\endalign
$$
Now $\hbar^{-1}G_{n\alpha}$ is an element of $\calG$. Since $W$
is assumed to be in $\exp\;\calG_{\le -1}$, the conjugation
$W*\hbar^{-1}G_{n\alpha}*W^{-1}$ again belongs to $\calG$.
Therefore $\hbar^{-1}B_{n\alpha}$ becomes an element of
$\calG_{\ge 0}$. Thus the consistency of the above equations
is ensured.

These flows are highly nonlinear and nontrivial in general, but
lowest flows contain trivial ones. For instance, the Zakharov-Shabat
operators for $(n,\alpha) = (0,0), (1,0), (0,1)$ are given by
$$
  B_{00} = 1, \quad B_{10} = \hbar\rd_x, \quad B_{01} = y.  \tag\eq
$$
Therefore
$$
  \dfrac{\rd W}{\rd t_{00}} = 0, \quad
  \dfrac{\rd W}{\rd t_{10}} = \dfrac{\rd W}{\rd x}, \quad
  \dfrac{\rd W}{\rd t_{01}} = \dfrac{\rd W}{\rd z},
                                                      \tag\eq
$$
i.e.,  $W$ is independent of $t_{00}$, and the flows of
$t_{10}$ and $t_{01}$ are just translations in $x$ and $z$.

The factorization relation, too, can be formulated in the
same form. The exponential operator $E(t)$ is now given by
$$
  E(t) = \exp_{*} \hbar^{-1} \Bigl(
          \sum_{n,\alpha} t_{n\alpha}(\hbar\rd_x)^n y^\alpha\Bigr),
                                                      \tag\eq
$$
where $\exp_{*}$ denotes the star exponential,
$$
  \exp_{*} A
    = 1 + \sum_{n=1}^\infty \frac{1}{n!} A*\cdots*A
                    \  (\text{$n$-fold star product}).
                                                      \tag\eq
$$
With this star exponential operator, the flows on the space
$\exp\;\calG_{\le -1}$ of dressing operators can be identified
with the action of $E(t)$ on the coset space
$\exp\;\calG/\exp\;\calG_{\ge 0}$.

\section{Lax and zero-curvature equations}

We now consider the operators
$$
  L = W * \hbar\rd_x * W^{-1}, \quad
  U = W * y * W^{-1}
                                                    \tag\eq
$$
as analogues of the Lax operators $L$ and $U_\alpha$ of the
$N$-component KP hierarchy. They can be written
$$
  L = \hbar\rd_z + \sum_{n=1}^\infty g_{n+1}(\hbar\rd_x)^{-n},
  \quad
  U = y + \sum_{n=1}^\infty u_n * L^{-n},
                                                     \tag\eq
$$
and the coefficients turn out to have smooth limit as $\hbar \to 0$,
$$
  g_n = g_n^{(0)}(t,x,y,z) + O(\hbar), \quad
  u_n = u_n^{(0)}(t,x,y,z) + O(\hbar).
                                        \tag\eqname\LimitOfGU
$$
[Proof: The dressing operator $W$ is assumed to be the star
product exponential of an element of $\calG_{-1}$. Evaluating
$W*\hbar\rd_x*W^{-1}$ and $W*y*W^{-1}$ by use of the general
formula
$$
    e^X Y e^{-X} = Y + [X,Y] + \frac{1}{2}[X,[X,Y]] + \cdots,
                                   \tag\eqname\AdjointFormula
$$
one can indeed prove the above fact. Q.E.D.]

Zakharov-Shabat operators are given by
$$
    B_{n\alpha} = \left( L^n * U^\alpha \right)_{\ge 0},
                                                      \tag\eq
$$
where we abbreviate $L^n = L * \cdots * L$ ($n$-fold star product)
and $U^\alpha = U * \cdots * U$ ($\alpha$-fold star product).

As a consequence of (\EqOfWTNA), these operators turn out
to satisfy the Lax equations
$$
  \hbar\dfrac{\rd L}{\rd t_{n\alpha}} = [ B_{n\alpha}, L ],
  \quad
  \hbar\dfrac{\rd U}{\rd t_{n\alpha}} = [ B_{n\alpha}, U ]
                                                   \tag\eq
$$
and the zero-curvature equations
$$
  \hbar\dfrac{\rd B_{m\alpha}}{\rd t_{n\beta}}
  - \hbar\dfrac{\rd B_{n\beta}}{\rd t_{m\alpha}}
  + [ B_{m\alpha}, B_{n\beta} ] = 0
                                                   \tag\eq
$$
with respect to star product commutator $ [A,B] = A*B -B*A $.

\section{Extended Lax equations}

We now introduce another pair of Lax operators. This is inspired
by the treatment of quasi-classical limit of the KP hierarchy
[\REF\TTdKP{
  Takasaki, K., and Takebe, T.,
  SDiff(2) KP hierarchy,
  in: A. Tsuchiya et al. (eds.),
  {\it Infinite Analysis\/}, RIMS Research Project 1991,
  Int. J. Mod. Phys. A7, Suppl. 1
  (World Scientific, Singapore, 1992).
}\TTdKP].
The extra Lax operators are given by
$$
\align
  &  M = W * E(t) * x * E(t)^{-1} * W^{-1},              \\
  &  V = W * E(t) * z * E(t)^{-1} * W^{-1},
                                           \tag\eq       \\
\endalign
$$
and satisfy the same Lax equations
$$
 \hbar\dfrac{\rd M}{\rd t_{n\alpha}} = [ B_{n\alpha}, M ],
 \quad
 \hbar\dfrac{\rd V}{\rd t_{n\alpha}} = [ B_{n\alpha}, V ]
                                                   \tag\eq
$$
as well as the canonical commutation relations
$$
\align
  & [L,M] = \hbar, \quad [U,V] = \hbar,                \\
  & \text{other commutators} = 0.              \tag\eq \\
\endalign
$$

Let us specify the structure of these extra Lax operators
in more detail. By use of formula (\AdjointFormula), one can
calculate the ``pre-dressing" by $E(t)$:
$$
\align
  &  E(t) * x * E(t)^{-1}
   = \sum_{n,\alpha} n t_{n\alpha}(\hbar\rd_x)^{n-1} y^\alpha
     + x,                                                  \\
  &  E(t) * z * E(t)^{-1}
   = \sum_{n,\alpha} \alpha t_{n\alpha} (\hbar\rd_x)^n y^{\alpha-1}
     + z.
                                                  \tag\eq  \\
\endalign
$$
Dressing by $W$ can easily be evaluated by the formulas
$$
\align
  & W*x*W^{-1} = x + (\text{element of }\calG_{\le -1}),   \\
  & W*z*W^{-1} = z + (\text{element of }\calG_{\le -1}),   \\
  & W*(\hbar\rd_x)^n y^\alpha * W^{-1} = L^n * U^\alpha.
                                                  \tag\eq \\
\endalign
$$
Thus, eventually, $M$ and $V$ turn out to be written
$$
\align
  & M = \sum_{n,\alpha} n t_{n\alpha} L^{n-1} * U^\alpha
        + x + \sum_{n=1}^\infty h_{n+1} * L^{-n-1},
                                                           \\
  & V = \sum_{n,\alpha} \alpha t_{n\alpha} L^n * U^{\alpha-1}
        + z + \sum_{n=1}^\infty v_n * L^{-n},
                                                  \tag\eq  \\
\endalign
$$
and the coefficients $h_n$ and $v_n$, like those of of $L$
and $U$, have smooth limit as $\hbar \to 0$:
$$
    h_n = h_n^{(0)}(t,x,y,z) + O(\hbar), \quad
    v_n = v_n^{(0)}(t,x,y,z) + O(\hbar).
                                       \tag\eqname\LimitOfHV
$$

Remark:
We could have defined $M = W*x*W^{-1}$ and $V = W*z*W^{-1}$,
but then extra terms emerge on the right hand side of the Lax
equations.  Exactly the same phenomenon takes place in the
KP hierarchy [\TTdKP]; the above definition of $M$ and $V$
mimics the formulation developed therein.  In this respect,
we should define $L$ and $U$, too, via the same ``pre-dressing"
by $E(t)$. This however gives the same result because $\hbar\rd_x$
and $y$ commute with $E(t)$.

\chapter{Hierarchy of noncommuting flows}

\noindent
By ``noncommuting flows" we mean a set of flows with generators
$G_i$ that do not necessary commute with each other. In general,
such noncommutativity causes serious troubles in formulating
a Lax formalism. In the following case, however, the generators
obey rather special commutation relations, so that we can write
down Lax equations etc. in almost the same explicit way as in
the case of commuting flows.

\section{Equation of dressing operator}

The second hierarchy consists of flows with three series of
time variables, $t = (t_1,t_2,\ldots)$, $p = (p_1,p_2,\ldots)$,
and $q = (q_1,q_2,\ldots)$.  The flows are generated by the
left action of the star product exponential
$$
    E(t,p,q) = \exp_{*} \hbar^{-1} \Bigl(
       \sum_{n=1}^\infty t_n (\hbar\rd)^n
       - \sum_{n=1}^\infty p_n z (\hbar\rd_x)^n
       + \sum_{n=1}^\infty q_n y (\hbar\rd_x)^n  \Bigr)
                                                   \tag\eq
$$
on the coset space $\exp\; \calG / \exp\;\calG_{\ge 0}$.
These flows can be converted into flows on the space
$\exp\;\calG_{\le -1}$ of dressing operators by the
factorization relation
$$
    W(t,p,q) * E(t,p,q) * W(0,0,0)^{-1}
    = \What(t,p,q) \in \exp \calG_{\ge 0}
                                                   \tag\eq
$$
connecting $W=W(t,p,q)$ with the initial value at
$(t,p,q) = (0,0,0)$. The generators of flows in
$E(t,p,q)$ are noncommutative,
$$
    \bigl[ \hbar^{-1}(\hbar\rd_x)^m y,
           \hbar^{-1}(\hbar\rd_x)^n z \bigr]
    = \hbar^{-1} (\hbar\rd_x)^{m+n},
                                                   \tag\eq
$$
but note that commutators are all central. Thus the algebraic
structure is very similar to Heisenberg algebras. This is a
key to the following calculations.

Because of this mild noncommutativity, one can explicitly
write down differential equations satisfied by $E(t,p,q)$ as:
$$
\align
  \hbar \dfrac{\rd E(t,p,q)}{\rd t_n}
  = & (\hbar\rd_x)^n E(t,p,q),
                                                        \\
  \hbar \dfrac{\rd E(t,p,q)}{\rd p_n}
  = & \Bigl( - z(\hbar\rd_x)^n
         - \frac{1}{2}\sum_{m=1}^\infty q_m(\hbar\rd_x)^{m+n}
      \Bigr) * E(t,p,q),
                                                        \\
  \hbar \dfrac{\rd E(t,p,q)}{\rd q_n}
  = & \Bigl(  y (\hbar\rd_x)^n
         + \frac{1}{2}\sum_{m=1}^\infty p_m(\hbar\rd_x)^{m+n}
      \Bigr) * E(t,p,q),
                                    \tag\eqname\EqOfETPQ  \\
\endalign
$$
[Proof: To derive these formulas, let us factorize $E(t,p,q)$
in two different ways:
$$
\align
  E(t,p,q)
  = & \exp \hbar^{-1} \Bigl(
               \sum_{n=1}^\infty t_n(\hbar\rd_x)^n
             - \sum_{n=1}^\infty p_n z (\hbar\rd_x)^n         \\
    &        - \frac{1}{2}\sum_{m,n=1}^\infty
                  p_m q_n (\hbar\rd_x)^{m+n}
      \Bigl)
      * \exp \hbar^{-1} \Bigl(
               \sum_{n=1}^\infty q_n y (\hbar\rd_x)^n \Bigr), \\
  = & \exp \hbar^{-1}\Bigl(
               \sum_{n=1}^\infty t_n(\hbar\rd_x)^n
             + \sum_{n=1}^\infty q_n y (\hbar\rd_x)^n         \\
    &        + \frac{1}{2}\sum_{m,n=1}^\infty
                  p_m q_n (\hbar\rd_x)^{m+n}
      \Bigr)
      * \exp \hbar^{-1} \Bigl(
             - \sum_{n=1}^\infty p_n z (\hbar\rd_x)^n \Bigr).
                                                       \tag\eq \\
\endalign
$$
Here we have used the Campbell-Hausdorff formula
$$
    e^X e^Y = \exp( X + Y + \frac{1}{2}[X,Y] + \cdots)
                                \tag\eqname\CampbellHausdorff
$$
for star product. (Note that because of the above commutation
relations of Heisenberg-type, multiple commutators disappear.)
The first factorized form of $E(t,p,q)$ is suited for calculating
$p$-derivatives, because $p_n$'s are included only in the first
factor, and all terms included therein commute with each other.
The second formula of (\EqOfETPQ) can thus be proven. Similarly,
the second factorized form of $E(t,p,q)$ can be used to evaluate
$q$-derivatives.  For $t$-derivatives, nothing subtle is required.
Q.E.D.]

Having these differential equations for $E(t,p,q)$, one can
derive differential equations satisfied by the dressing
operator as:
$$
\align
  \hbar \dfrac{\rd W}{\rd t_n} * W^{-1}
  = & - \Bigl( W * (\hbar\rd_x)^n * W^{-1} \Bigr)_{\le -1},
                                                             \\
  \hbar \dfrac{\rd W}{\rd p_n} * W^{-1}
  = & - \Bigl( - W * z(\hbar\rd_n)^n * W^{-1}
               - \frac{1}{2} \sum_{m=1}^\infty
           q_m W * (\hbar\rd_x)^{m+n} * W^{-1} \Bigr)_{\le -1},
                                                             \\
  \hbar \dfrac{\rd W}{\rd q_n} * W^{-1}
  = & - \Bigl(   W * y(\hbar\rd_x)^n * W^{-1}
               + \frac{1}{2} \sum_{m=1}^\infty
           p_m W * (\hbar\rd_x)^{m+n} * W^{-1} \Bigr)_{\le -1}.
                                      \tag\eqname\EqOfWTPQ   \\
\endalign
$$
[Proof: We simply repeat the reasoning in the end of Section 2.
For instance, let us check the second equation. Differentiating
the factorization relation and eliminating the initial value
$W(0,0,0)$, one obtains the basic relation
$$
  \left( \dfrac{\rd W}{\rd p_n}
  + W * \dfrac{\rd E}{\rd p_n} * E^{-1} \right) * W^{-1}
  = \dfrac{\rd \What}{\rd p_n} * \What^{-1}.
                                                  \tag\eq
$$
Since the right hand side should be an element of
$\calG_{\ge 0}$,
$$
  \Bigl(\text{left hand side of }(\?) \Bigr)_{\le -1} = 0.
                                                  \tag\eq
$$
By use of (\EqOfETPQ), one can easily check that this gives
the second equation of (\EqOfWTPQ). The other equations can
be proven in the same manner. Q.E.D.]

\section{Lax and zero-curvature equations}

We now define Lax operators just like the definition of
$M$ and $V$ in the commuting hierarchy:
$$
\align
    L =& W * E(t,p,q) * \hbar\rd_x * E(t,p,q)^{-1} * W^{-1},  \\
    M =& W * E(t,p,q) * x * E(t,p,q)^{-1} * W^{-1},           \\
    U =& W * E(t,p,q) * y * E(t,p,q)^{-1} * W^{-1},           \\
    V =& W * E(t,p,q) * z * E(t,p,q)^{-1} * W^{-1}.
                                                      \tag\eq \\
\endalign
$$
``Pre-dressing" by $E(t,p,q)$ can be calculated by use of
(\AdjointFormula) as:
$$
\align
  &  E(t,p,q) * \hbar\rd_x * E(t,p,q)^{-1} = \hbar\rd_x,      \\
  &  E(t,p,q) * x * E(t,p,q)^{-1}
     =  \sum_{n=1}^\infty n t_n (\hbar\rd_x)^{n-1}
        - \sum_{n=1}^\infty n p_n z (\hbar\rd_x)^{n-1}        \\
  &  \quad\quad
                + \sum_{n=1}^\infty n q_n y (\hbar\rd_x)^{n-1}
                - \frac{1}{2} \sum_{m,n=1}^\infty
                           (m-n) p_m q_n (\hbar\rd_x)^{m+n-1}
                + x,                                          \\
  &  E(t,p,q) * y * E(t,p,q)^{-1}
     = \sum_{n=1}^\infty p_n (\hbar\rd_x)^n + y,              \\
  &  E(t,p,q) * z * E(t,p,q)^{-1}
     = \sum_{n=1}^\infty q_n (\hbar\rd_x)^n + z,
                                                     \tag\eq  \\
\endalign
$$
Dressing by $W$, then, yields the following expression of the Lax
operators.
$$
\align
  L =& \hbar\rd_x + \sum_{n=1}^\infty g_{n+1}(\hbar\rd_x)^{-n},
                                                                  \\
  M =& \sum_{n=1}^\infty n t_n L^{n-1}
       - \sum_{n=1}^\infty n p_n V * L^{n-1}
       + \sum_{n=1}^\infty n q_n U * L^{n-1}                      \\
     & + \frac{1}{2} \sum_{m,n=1}^\infty (m-n) p_m q_n L^{m+n-1}
       + x + \sum_{n=1}^\infty h_{n+1} * L^{-n-1},
                                                                  \\
  U =& \sum_{n=1}^\infty p_n L^n + y
       + \sum_{n=1}^\infty u_n * L^{-n},
                                                                  \\
  V =& \sum_{n=1}^\infty q_n L^n + z
       + \sum_{n=1}^\infty v_n * L^{-n}.
                                                         \tag\eq  \\
\endalign
$$
Just as in the commuting hierarchy, cf. (\LimitOfGU) and (\LimitOfHV),
the coefficients on the right hand sides have smooth limit as
$\hbar \to 0$.

Let us now write down differential equations satisfied by these
Lax operators. First note that (\EqOfWTPQ) can now be written
$$
\align
   \hbar \dfrac{\rd W}{\rd t_n}
    =& - \Bigl( L^n  \Bigr)_{\le -1} * W,                     \\
   \hbar \dfrac{\rd W}{\rd p_n}
    =& - \Bigl( -V*L^n + \frac{1}{2}\sum_{m=1}^\infty
                            q_m L^{m+n} \Bigr)_{\le -1} * W,  \\
   \hbar \dfrac{\rd W}{\rd q_n}
    =& - \Bigl(  U*L^n - \frac{1}{2}\sum_{m=1}^\infty
                            p_m L^{m+n} \Bigr)_{\le -1} * W.
                                                    \tag\eq   \\
\endalign
$$
Remarkably, $M$ does not take place on the right hand side.
This implies that one does not actually need $M$ for defining
Zakharov-Shabat operators. One can indeed derive, from
the above equations, the Lax equations
$$
\align
  & \hbar \dfrac{\rd L}{\rd t_n} = [B_n, L],  \quad
    \hbar \dfrac{\rd L}{\rd p_n} = [C_n, L],  \quad
    \hbar \dfrac{\rd L}{\rd q_n} = [D_n, L],                 \\
  & \hbar \dfrac{\rd U}{\rd t_n} = [B_n, U],  \quad
    \hbar \dfrac{\rd U}{\rd p_n} = [C_n, U],  \quad
    \hbar \dfrac{\rd U}{\rd q_n} = [D_n, U],                 \\
  & \hbar \dfrac{\rd V}{\rd t_n} = [B_n, V],  \quad
    \hbar \dfrac{\rd V}{\rd p_n} = [C_n, V],  \quad
    \hbar \dfrac{\rd V}{\rd q_n} = [D_n, V].
                                                      \tag\eq \\
\endalign
$$
The Zakharov-Shabat operators $B_n$, $C_n$ and $D_n$ are
given by
$$
\align
  B_n =& \Bigl( L^n \Bigr)_{\ge 0},                          \\
  C_n =& \Bigl( - V*L^n + \frac{1}{2}\sum_{m=1}^\infty
                           q_m L^{m+n} \Bigr)_{\ge 0},       \\
  D_n =& \Bigl(   U*L^n - \frac{1}{2}\sum_{m=1}^\infty
                           p_m L^{m+n} \Bigr)_{\ge 0},
                                                     \tag\eq \\
\endalign
$$
and satisfy the zero-curvature equations
$$
\align
    \hbar \dfrac{\rd B_m}{\rd t_n}
    - \hbar \dfrac{\rd B_n}{\rd t_m}
    + [ B_m, B_n ] =& 0,
                                                            \\
    \hbar \dfrac{\rd B_m}{\rd p_n}
    - \hbar \dfrac{\rd C_n}{\rd t_m}
    + [ B_m, C_n ] =& 0,
                                                            \\
    \hbar \dfrac{\rd B_m}{\rd q_n}
    - \hbar \dfrac{\rd D_n}{\rd t_m}
    + [ B_m, D_n ] =& 0,
                                                            \\
    \hbar \dfrac{\rd C_m}{\rd p_n}
    - \hbar \dfrac{\rd C_n}{\rd p_m}
    + [ C_m, C_n ] =& 0,
                                                            \\
    \hbar \dfrac{\rd C_m}{\rd q_n}
    - \hbar \dfrac{\rd D_n}{\rd p_m}
    + [ C_m, D_n ] =& 0,
                                                            \\
    \hbar \dfrac{\rd D_m}{\rd q_n}
    - \hbar \dfrac{\rd D_n}{\rd q_m}
    + [ D_m, D_n ] =& 0.
                                                  \tag\eq   \\
\endalign
$$
Thus the Lax and zero-curvature equations are closed within
$L$, $U$ and $V$. (Of course $M$, too, satisfies the same
Lax equations as the other three Lax operators.)  Let us focus
the following consideration on these three Lax operators.
Note that they obey the commutation relations
$$
   [L,U] = [L,V] = 0, \quad [U,V] = \hbar.              \tag\eq
$$

\chapter{Reduction to selfdual gravity}

\noindent
Moyal algebraic deformations of selfdual gravity emerge from
the hierarchy of $(t,p,q)$ flows as a kind of ``dimensional
reduction''.  Actually, connection to selfdual gravity takes two
distinct forms, which correspond to two different local expressions
of selfdual gravity --- the first and second heavenly equations
of Plebanski
[\REF\PlSDG{
  Plebanski, J.F.,
  Some solutions of complex Einstein equations,
  J. Math. Phys. 16 (1975), 2395-2402.
}\PlSDG].

\section{Constraints and reduced variables}

The reduction of the $(t,p,q)$ flows is defined by the
constraints
$$
   \dfrac{\rd w_n}{\rd x} = 0,  \quad n=1,2,\ldots
                                                   \tag\eq
$$
on the coefficients of the dressing operator.  This forces
$L$ and $B_n$ to be trivial,
$$
    L = \hbar\rd_x, \quad
    B_n = (\hbar\rd_x)^n.
                                                   \tag\eq
$$
In particular,
$$
    \dfrac{\rd W}{\rd t_n} = 0, \quad
    \dfrac{\rd L}{\rd t_n} = \dfrac{\rd U}{\rd t_n}
    = \dfrac{\rd V}{\rd t_n} = 0.
                                                    \tag\eq
$$

Under these constraints, $\hbar\rd_x$ may be replaced by a
``spectral parameter" $\lambda$.  The dressing operator then
turns into a Laurent series with Moyal algebraic coefficients:
$$
  W(\lambda)
  = 1 + \sum_{n=1}^\infty w_n(\hbar,t,p,q,y,z)\lambda^{-n}.
                                                   \tag\eq
$$
This type of constraints are also used in the theory of
multi-component KP hierarchies [\KPGen] to derive an
$N$-component version of the AKNS (Ablowitz-Kaup-Newell-Segur)
or ZS (Zakharov-Shabat) hierarchy. The role of dressing
operators is played therein by a Laurent series with matrix
coefficients:
$$
  W(\lambda) = 1 + \sum_{n=1}^\infty w_n \lambda^{-n}, \quad
  w_n(t) \in \gl(N).
                                                   \tag\eq
$$

The Lax and Zakharov-Shabat operators, too, can be represented
by Laurent series. As noted above, $L$ is now trivial.
Nontrivial dynamical contents are carried by $U$ and $V$,
which are now replaced by Laurent series of the form
$$
\align
    U(\lambda) =& \sum_{n=1}^\infty p_n \lambda^n + y
                +\sum_{n=1}^\infty u_n \lambda^{-n},
                                                           \\
    V(\lambda) =& \sum_{n=1}^\infty q_n \lambda^n + z
                +\sum_{n=1}^\infty v_n \lambda^{-n}.
                                             \tag\eq       \\
\endalign
$$

Algebraic relations among the Lax operators are carried over to
these Laurent series. For instance, $U(\lambda)$ and $V(\lambda)$
obeys the canonical commutation relation
$$
   [ U(\lambda), V(\lambda) ] = \hbar
                                               \tag\eq
$$
and the dressing relations
$$
\align
   U(\lambda)
   =& W(\lambda) * \Bigl( \sum_{n=1}^\infty p_n \lambda^n
                                + y  \Bigr) * W(\lambda)^{-1}   \\
   =& \sum_{n=1}^\infty p_n \lambda^n
      + W(\lambda) * y * W(\lambda)^{-1},
                                                                \\
   V(\lambda)
   =& W(\lambda) * \Bigl( \sum_{n=1}^\infty q_n \lambda^n
                                + z  \Bigr) * W(\lambda)^{-1}   \\
   =& \sum_{n=1}^\infty q_n \lambda^n
      + W(\lambda) * z * W(\lambda)^{-1}.
                                                     \tag\eq    \\
\endalign
$$
Laurent series corresponding to Zakharov-Shabat operators
will be given in the next subsection.

\section{Equations of flows in terms of Laurent series}

The factorization relation connecting $W$ and its initial value
is now translated into a factorization relation for
$W(\lambda) = W(p,q,\lambda)$ and its initial value at $p=q=0$:
$$
    W(p,q,\lambda) * E(p,q,\lambda) * W(0,0,\lambda)^{-1}
    = \What(p,q,\lambda),
                                        \tag\eqname\FactorWLam
$$
where $\What(p,q,\lambda)$ is a Laurent series to be obtained
from $\What$ by replacing $\hbar\rd_x \to \lambda$ (which, too,
turns out to be independent of $x$ and $t$), and $E(p,q,\lambda)$
is the Laurent series
$$
    E(p,q,\lambda) = \exp_{*} \hbar^{-1} \Bigl(
        - \sum_{n=1}^\infty p_n z \lambda^n
        + \sum_{n=1}^\infty q_n y \lambda^n \Bigr).
                                                    \tag\eq
$$
(We put $t=0$ here and in the following.) The Lie algebra $\calG$
of pseudo-differential operators should thus be redefined to be a
Lie algebra of Laurent series with Moyal algebraic coefficients:
$$
\align
   \calG       =& \{ A \mid A = \sum a_n(\hbar,y,z)\lambda^n \}
         = \calG_{\ge 0} \oplus \calG_{\le -1},
                                                             \\
   (\quad)_{+} =& (\quad)_{\ge 0}: \
       \text{projection onto }\lambda^0, \lambda^1,\cdots,
                                                             \\
   (\quad)_{-} =& (\quad)_{\le -1}: \
       \text{projection onto }\lambda^{-1},\lambda^{-2},\cdots.
                                                    \tag\eq  \\
\endalign
$$

Differential equations for the Lax and dressing operators, too,
can be translated into the language of Laurent series. First,
$W(\lambda)$ satisfy the equations
$$
\align
  \hbar \dfrac{\rd W(\lambda)}{\rd p_n}
  = &   \left( V(\lambda)\lambda^n \right)_{\le -1} * W(\lambda),
                                                               \\
  \hbar \dfrac{\rd W(\lambda)}{\rd q_n}
  = & - \left( U(\lambda)\lambda^n \right)_{\le -1} * W(\lambda).
                                          \tag\eqname\EqOfWLamPQ
                                                                \\
\endalign
$$
Similarly, $U(\lambda)$ and $V(\lambda)$ satisfy the Lax equations
$$
\align
  & \hbar \dfrac{\rd U(\lambda)}{\rd p_n}
    = [C_n(\lambda),U(\lambda)]
    = [-\left( V(\lambda)\lambda^n \right)_{\ge 0}, U(\lambda)],
                                                            \\
  & \hbar \dfrac{\rd U(\lambda)}{\rd q_n}
    = [D_n(\lambda),U(\lambda)]
    = [ \left( U(\lambda)\lambda^n \right)_{\ge 0}, U(\lambda)],
                                                            \\
  & \hbar \dfrac{\rd V(\lambda)}{\rd p_n}
    = [C_n(\lambda),V(\lambda)]
    = [-\left( V(\lambda)\lambda^n \right)_{\ge 0}, V(\lambda)],
                                                            \\
  & \hbar \dfrac{\rd V(\lambda)}{\rd q_n}
    = [D_n(\lambda),V(\lambda)]
    = [ \left( U(\lambda)\lambda^n \right)_{\ge 0}, V(\lambda)],
                                        \tag\eqname\EqOfUVLam
                                                            \\
\endalign
$$
where $C_n(\lambda)$ and $D_n(\lambda)$ are given by
$$
\align
  C_n(\lambda) =& - \left( V(\lambda)\lambda^n \right)_{\ge 0}
                + \frac{1}{2}\sum_{m=1}^\infty q_m \lambda^{m+n},
                                                            \\
  D_n(\lambda) =&   \left( U(\lambda)\lambda^n \right)_{\ge 0}
                - \frac{1}{2}\sum_{m=1}^\infty p_m \lambda^{m+n}
                                                     \tag\eq
                                                            \\
\endalign
$$
and obey the zero-curvature equations
$$
\align
  &   \hbar\dfrac{\rd C_m(\lambda)}{\rd p_n}
    - \hbar\dfrac{\rd C_n(\lambda)}{\rd p_m}
    + [C_m(\lambda),C_n(\lambda)] = 0,                       \\
  &   \hbar\dfrac{\rd C_m(\lambda)}{\rd q_n}
    - \hbar\dfrac{\rd D_n(\lambda)}{\rd p_m}
    + [C_m(\lambda),D_n(\lambda)] = 0,                       \\
  &   \hbar\dfrac{\rd D_m(\lambda)}{\rd q_n}
    - \hbar\dfrac{\rd D_n(\lambda)}{\rd q_m}
    + [D_m(\lambda),D_n(\lambda)] = 0.
                                      \tag\eqname\EqOfCDLam  \\
\endalign
$$

\section{Relation to second heavenly equation}

We now show that the above reduced hierarchy is related to
a Moyal algebraic deformation of selfdual gravity. This
deformation corresponds to Plebanski's ``second heavenly
equation" [\PlSDG]:
$$
    \frac{\rd^2 \Theta}{\rd y \rd q}
    - \frac{\rd^2 \Theta}{\rd z \rd p}
    + \left\{ \frac{\rd \Theta}{\rd y},
             \frac{\rd \Theta}{\rd z} \right\} = 0.
                                      \tag\eqname\SecondHeaven
$$
Strachan's equation [\StMoyalSDG], on the other hand, is a
deformation of Plebanski's ``first heavenly equation" [\PlSDG]:
$$
    \Bigl\{ \frac{\rd\Omega}{\rd p},
            \frac{\rd\Omega}{\rd q}
    \Bigr\}\widehat{\vphantom{\Bigr\}}}
    =  \frac{\rd^2 \Omega}{\rd p \rd \phat}
         \frac{\rd^2 \Omega}{\rd q \rd \qhat}
     - \frac{\rd^2 \Omega}{\rd p \rd \qhat}
        \frac{\rd^2 \Omega}{\rd q \rd \phat} = 1,
                                      \tag\eqname\FirstHeaven
$$
where $(p,q,\phat,\qhat)$ are another set of variables, and
$\{\quad,\quad\}\widehat{\vphantom{\}}}\;$ stands for the Poisson
bracket in $(\phat,\qhat)$. Comparing the above hierarchy with
hierarchies constructed in Ref. \TaSDG , and identifying
$$
    p_1 = p, \quad q_1 = q,              \tag\eqname\POneQOne
$$
one will see that the above hierarchy gives a Moyal algebraic
deformation of a hierarchy constructed therein for the second
heavenly equation. Furthermore, Plebanski's $\Theta$ potential
turns out to be linked with the next-to-leading term $w_1$ of
$W(\lambda)$:
$$
    \Theta = - \hbar w_1.
                                      \tag\eqname\ThetaIsWLamOne
$$

Let us show how to deduce these facts.  To this end, we rewrite
(\EqOfWLamPQ) as
$$
\align
  \hbar \dfrac{\rd W(\lambda)}{\rd p_n} * W(\lambda)^{-1}
  = &   \left( V(\lambda)\lambda^n \right)_{\le -1},
                                                          \\
  \hbar \dfrac{\rd W(\lambda)}{\rd q_n} * W(\lambda)^{-1}
  = & - \left( U(\lambda)\lambda^n \right)_{\le -1},
                                            \tag\eq       \\
\endalign
$$
and extract $\lambda^{-1}$ terms of both hand sides. This
results in the relations
$$
  \hbar \dfrac{\rd w_1}{\rd p_n} = v_{n+1}, \quad
  \hbar \dfrac{\rd w_1}{\rd q_n} = - u_{n+1}.
                                               \tag\eq
$$
[Actually, these relations can be extended to the case of
$n=0$ by identifying
$$
    p_0 = y, \quad q_0 = z,
                                    \tag\eqname\PZeroQZero
$$
as one can directly check from the definition of $U(\lambda)$
and $V(\lambda)$.] In terms of the function $\Theta$ defined by
(\ThetaIsWLamOne), these relations can be rewritten
$$
    u_n = \dfrac{\rd \Theta}{\rd q_{n-1}}, \quad
    v_n = - \dfrac{\rd \Theta}{\rd p_{n-1}}.
                                     \tag\eqname\UnVnByTheta
$$
In fact, exactly the same relations for selfdual gravity have
been derived in Ref. \TaSDG .  Furthermore, repeating technical
calculations presented therein, one can show that Eqs.
(\EqOfCDLam) are actually equivalent to:
$$
\align
  & \dfrac{\rd v_m}{\rd p_n} - \dfrac{\rd v_n}{\rd p_m}
  + \{ v_n, v_m \}_\hbar = 0,                                 \\
  & \dfrac{\rd u_m}{\rd p_n} + \dfrac{\rd v_n}{\rd q_m}
  + \{ v_n, u_m \}_\hbar = 0,                                 \\
  & \dfrac{\rd u_m}{\rd p_n} - \dfrac{\rd u_n}{\rd p_m}
  - \{ u_n, u_m \}_\hbar = 0,
    \quad m,n=0,1,\ldots.
                                                      \tag\eq
                                                              \\
\endalign
$$
By (\UnVnByTheta), these equations can be rewritten
$$
\align
  &  \dfrac{\rd^2 \Theta}{\rd p_n \rd q_{m-1}}
   - \dfrac{\rd^2 \Theta}{\rd p_m \rd q_{n-1}}
   - \left\{ \dfrac{\rd \Theta}{\rd p_{n-1}},
             \dfrac{\rd \Theta}{\rd p_{m-1}} \right\}_\hbar = 0,
                                                              \\
  &  \dfrac{\rd^2 \Theta}{\rd p_n \rd q_{m-1}}
   - \dfrac{\rd^2 \Theta}{\rd q_m \rd p_{n-1}}
   - \left\{ \dfrac{\rd \Theta}{\rd p_{n-1}},
             \dfrac{\rd \Theta}{\rd q_{m-1}} \right\}_\hbar = 0,
                                                              \\
  &  \dfrac{\rd^2 \Theta}{\rd q_n \rd q_{m-1}}
   - \dfrac{\rd^2 \Theta}{\rd q_m \rd q_{n-1}}
   - \left\{ \dfrac{\rd \Theta}{\rd q_{n-1}},
             \dfrac{\rd \Theta}{\rd q_{m-1}} \right\}_\hbar = 0.
                                                       \tag\eq
                                                              \\
\endalign
$$
The second equation with $m=n=1$ gives a Moyal algebraic
deformation of second heavenly equation (\SecondHeaven).
In the limit of $\hbar \to 0$, the Moyal bracket
$\{\quad,\quad\}_\hbar$ turns into the Poisson bracket
$\{\quad,\quad\}$, and these equations reproduce a
hierarchy constructed in Ref. \TaSDG .

\section{Relation to first heavenly equation}

Relation to the first heavenly equation is more involved.
First heavenly equation (\FirstHeaven) is written in terms of
the Poisson bracket $\{\quad,\quad\}\widehat{\vphantom{\}}}\;$
in $(\phat,\qhat)$, and Strachan's idea is to deform this
bracket into a Moyal bracket as:
$$
    \Bigl\{ \frac{\rd\Omega}{\rd p},
            \frac{\rd\Omega}{\rd q}
            \Bigr\}_\hbar\widehat{\vphantom{\Bigr\}}}
    =  \frac{\rd^2 \Omega}{\rd p \rd \phat}
         \frac{\rd^2 \Omega}{\rd q \rd \qhat}
     - \frac{\rd^2 \Omega}{\rd p \rd \qhat}
        \frac{\rd^2 \Omega}{\rd q \rd \phat}
     + O(\hbar)
    = 1.
                             \tag\eqname\MoyalFirstHeaven
$$
These brackets differ from our brackets in $(y,z)$. Nevertheless,
very curiously and remarkably, it turns out that a similar Moyal
algebraic analogue of the first heavenly equation emerges in a
``gauge transformation" of the present hierarchy.

Let us first specify this gauge transformation. By
``gauge transformations" we mean symmetries of Eqs.
(\EqOfWLamPQ-\EqOfCDLam) given by
$$
\align
  W(\lambda) & \to W^g(\lambda) = g^{-1} * W(\lambda),             \\
  U(\lambda) & \to U^g(\lambda) = g^{-1} * U(\lambda) * g,         \\
  V(\lambda) & \to V^g(\lambda) = g^{-1} * V(\lambda) * g,         \\
  C_n(\lambda) & \to C_n^g(\lambda)
    = g^{-1} * C_n(\lambda) * g - g^{-1} * \dfrac{\rd g}{\rd p_n}, \\
  D_n(\lambda) & \to D_n^g(\lambda)
    = g^{-1} * D_n(\lambda) * g - g^{-1} *\dfrac{\rd g}{\rd q_n},
                                                              \tag\eq
                                                                   \\
\endalign
$$
where $g$ is a function of $(y,z,p,q)$ and independent of
$\lambda$.  With suitable choice of $g$ (i.e., ``a gauge-fixing"),
$\lambda^0$ terms in $C_n^g(\lambda)$ and $D_n(\lambda)$ can
be eliminated. Such a gauge-fixing is realized by
$$
    g = \what_0,                                           \tag\eq
$$
where $\what_0$ is the leading term of the Laurent series
$\What(\lambda)$ that arises in the factorization relation:
$$
    \What(\lambda) = \what_0 + \what_1 \lambda^1 + \cdots.  \tag\eq
$$

The factorization relation, too, is gauge-covariant
if $\What(\lambda)$ is also transformed as
$$
  \What(\lambda) \to \What^g(\lambda) = g^{-1} * \What(\lambda).
                                                            \tag\eq
$$
In the language of the factorization relation, the above special
gauge transformation simply means changing a normalization
condition of solutions.  By the above gauge transformation
(note that $g|_{p=q=0} = 1$), indeed, the original normalization
condition
$$
    W(\lambda)\bigr|_{\lambda=\infty} = 1                   \tag\eq
$$
is converted into a new condition of the form
$$
    \What^g(\lambda)\bigr|_{\lambda=0} = 1.                 \tag\eq
$$
This corresponds to taking a new direct sum decomposition of $\calG$:
$$
\align
        \calG =& \calG_{\ge 1} \oplus \calG_{\le 0},         \\
  (\quad)_{+} =& (\quad)_{\ge 0}: \
      \text{projection onto}\ \lambda^1, \lambda^2, \ldots,  \\
  (\quad)_{-} =& (\quad)_{\le 0}: \
      \text{projection onto}\ \lambda^0, \lambda^{-1}, \ldots,
                                                     \tag\eq \\
\endalign
$$
Repeating previous calculations in this new setting, one can
derive the following equations, which are gauge transformations
of (\EqOfWLamPQ).
$$
\align
  & \hbar \dfrac{\rd W^g(\lambda)}{\rd p_n}
  =    \Bigl( V^g(\lambda)\lambda^n \Bigr)_{\le 0} * W^g(\lambda),
                                                                   \\
  & \hbar \dfrac{\rd W^g(\lambda)}{\rd q_n}
  =  - \Bigl( U^g(\lambda)\lambda^n \Bigr)_{\le 0} * W^g(\lambda),
                                                         \tag\eq
                                                                   \\
\endalign
$$
This kind of gauge transformations are also known for the
KP hierarchy etc.
[\REF\KoKP{
  Konopelchenko, B.,
  Matrix Sato theory and integrable equations in 2+1 dimensions,
  in: M. Boiti et al. (eds.),
  {\it Nonlinear Evolution Equations and Dynamical Systems\/}
  (NEEDS '91)
  (World Scientific, Singapore, 1992).
}\KoKP].

Let us now return to the issue of selfdual gravity. We now
augue that the function
$$
    \Omega = - \hbar \what^g_1
                                 \tag\eqname\OmegaIsWhatOne
$$
satisfies a Moyal algebraic analogue of the first heavenly
equation,
$$
    \left\{ \dfrac{\rd \Omega}{\rd p},
            \dfrac{\rd \Omega}{\rd q} \right\}_\hbar
    =  \dfrac{\rd^2\Omega}{\rd p \rd y}
         \dfrac{\rd^2\Omega}{\rd q \rd z}
     - \dfrac{\rd^2\Omega}{\rd p \rd z}
         \dfrac{\rd^2~\Omega}{\rd q \rd y}
     + O(\hbar)
    = 1,
                              \tag\eqname\OurMoyalFirstHeaven
$$
where $(p_1,q_1)$ are identified with $(p,q)$ as in (\POneQOne),
and $\what^g_1$ is the next-to-leading term of Laurent
expansion of $\What^g(\lambda)$,
$$
    \What^g(\lambda) = 1 + \what^g_1 \lambda + \cdots.
                                                      \tag\eq
$$

Let us show details. As a consequence of the factorization
relation, $\What^g(\lambda)$ satisfies the same equation as
$W^g(\lambda)$:
$$
\align
  & \hbar \dfrac{\rd \What^g(\lambda)}{\rd p_n}
  =    \Bigl( V^g(\lambda)\lambda^n \Bigr)_{\le 0}
        * \What^g(\lambda),
                                                                \\
  & \hbar \dfrac{\rd \What^g(\lambda)}{\rd q_n}
  =  - \Bigl( U^g(\lambda)\lambda^n \Bigr)_{\le 0}
        * \What^g(\lambda).
                                                     \tag\eq    \\
\endalign
$$
Then imitating the derivation of (\UnVnByTheta), we can show that
$$
    u^g_n = - \dfrac{\rd \Omega}{\rd q_{n+1}}, \quad
    v^g_n =   \dfrac{\rd \Omega}{\rd p_{n+1}},
                                     \tag\eqname\UgnVgnByOmega
$$
where $u^g_n$ and $v^g_n$ are Laurent coefficients of
$U^g(\lambda)$ and $V^g(\lambda)$,
$$
\align
  U^g(\lambda) =& \sum_{n=1}^\infty p_n \lambda^n + u^g_0
                + \sum_{n=1}^\infty u^g_n \lambda^{-n},     \\
  V^g(\lambda) =& \sum_{n=1}^\infty q_n \lambda^n + v^g_0
                + \sum_{n=1}^\infty v^g_n \lambda^{-n}.
                                                    \tag\eq \\
\endalign
$$
In particular,
$$
    u^g_0 = -\dfrac{\rd\Omega}{\rd q_1}, \quad
    v^g_0 =  \dfrac{\rd\Omega}{\rd p_1}.
                                \tag\eqname\UgZeroVgZero
$$
On the other hand, by the construction, $u^g_0$ and $v^g_0$
are linked with the canonical conjugate pair $y$ and $z$ as
$$
    u^g_0 = g^{-1} * y * g, \quad
    v^g_0 = g^{-1} * z * g,                      \tag\eq
$$
hence canonical conjugate in themselves:
$$
   \{ u^g_0, v^g_0 \}_\hbar = 1.                 \tag\eq
$$
Substitution of (\UgZeroVgZero) into this relation yields
(\OurMoyalFirstHeaven).  Furthermore, as in the case of
$\Theta$, one can derive an infinite system of equations
for $\Omega$. Those equations give a Moyal algebraic
deformation of another hierarchy constructed in Ref. \TaSDG .

We are thus in a somewhat puzzled situation. Our Moyal
algebraic deformation (\OurMoyalFirstHeaven) has exactly the
same structure as Strachan's equation (\MoyalFirstHeaven),
but the Moyal brackets are living in apparently different
planes, i.e., $(y,z)$ and $(\phat,\qhat)$. In the standard
treatment of selfdual gravity, the two coordinate systems
$(p,q,y,z)$ and $(p,q,\phat,\qhat)$ are distinct and never
identical (unless the spacetime is flat).

This problem will be resolved in a Toda version of our Moyal
algebraic nonabelian KP hierarchies. The dressing operator
approach of Ref. \TaMoyalSDG\ , actually, employs two dressing
operators rather than a single one as we now considering.
This situation resembles the Toda hierarchy
[\REF\UTToda{
  Ueno, K., and Takasaki, K.,
  Toda lattice hierarchy,
  in: K. Okamoto (ed.),
  {\it Group Representations and Systems
  of Differential Equations},
  Advanced Studies in Pure Math. 4
  (North-Holland, Amsterdam, 1984).
}\UTToda],
which, too, is based on two dressing operators. (Actually,
although not clearly mentioned therein, fundamental ideas
in Ref. \TaMoyalSDG\ are rather borrowed from the theory of
the Toda hierarchy.) A natural framework in which to embed
Strachan's equation is thus a Moyal algebraic version
of the Toda hierarchy.

\chapter{Quasi-classical limit}

\noindent
The Moyal algebraic hierarchies of both commuting and noncommuting
types turn out to have quasi-classical ($\hbar \to 0$) limit.
Hierarchies in this limit, too, allow Lax- and zero-curvature-type
representations, but commutators in the ordinary framework are
now replaced by Poisson brackets. The situation is parallel to
the KP hierarchy; its quasi-classical limit is the dispersionless
KP hierarchy, and Poisson brackets are two dimensional. In the
present setting, Poisson brackets are four dimensional.

\section{Prescription of quasi-classical limit}

Let us recall the derivation of the dispersionless KP hierarchy
as quasi-classical limit of the ordinary KP hierarchy. The
canonical conjugate Lax operators $L$ and $M$ of the KP hierarchy
[\TTdKP], in the presence of $\hbar$, can be written
$$
\align
    L =& \hbar\rd_x
        + \sum_{n=1}^\infty g_{n+1} (\hbar\rd_x)^{-n},          \\
    M =& \sum_{n=1}^\infty n t_n L^{n-1} + x
        + \sum_{n=1}^\infty h_{n+1} L^{-n-1},
                                                        \tag\eq \\
\endalign
$$
and satisfy the canonical commutation relation
$$
    [ L, M ] = \hbar                                    \tag\eq
$$
and the Lax equations
$$
    \dfrac{\rd L}{\rd t_n} = [ B_n, L ], \quad
    \dfrac{\rd M}{\rd t_n} = [ B_n, M ],                \tag\eq
$$
where
$$
    B_n = \left( L^n \right)_{\ge 0}.                    \tag\eq \\
$$
The coefficients are assumed to behave smoothly as $\hbar \to 0$:
$$
    g_n = g^{(0)}_n(t,x) + O(\hbar), \quad
    h_n = h^{(0)}_n(t,x) + O(\hbar).                     \tag\eq
$$

In the limit of $\hbar \to 0$, the canonical conjugate pair
$(\hbar\rd_x,x)$ of operators are replaced by a canonical
conjugate pair $(k,x)$ of coordinates on a two dimensional
symplectic manifold.  The role of the Lax operators $L$
and $M$ are then played by the Laurent series
$$
\align
    \calL =& k + \sum_{n=1}^\infty g^{(0)}_{n+1} k^{-n},
                                                               \\
    \calM =& \sum_{n=1}^\infty n t_n \calL^{n-1} + x
             +\sum_{n=1}^\infty h^{(0)}_{n+1} \calL^{-n-1},
                                                          \tag\eq
\endalign
$$
which satisfies the canonical Poisson relation
$$
    \{ \calL, \calM \} = 1                              \tag\eq
$$
and the Lax-like equations
$$
    \dfrac{\rd \calL}{\rd t_n} = \{ \calB_n, \calL \}   \quad
    \dfrac{\rd \calM}{\rd t_n} = \{ \calB_n, \calM \}   \tag\eq
$$
with respect to the Poisson bracket
$$
    \{ A, B \} = \frac{\rd A}{\rd k} \frac{\rd B}{\rd x}
                -\frac{\rd A}{\rd x} \frac{\rd B}{\rd k}.
                                                        \tag\eq
$$
Here $\calB_n$'s are polynomials of the form
$$
    \calB_n = \left( \calL^n \right)_{\ge 0},           \tag\eq
$$
and the projection operators $(\quad)_{\ge 0}$ and
$(\quad)_{\le -1}$ are defined by
$$
\align
   (\quad)_{\ge 0}:& \
       \text{projection onto } k^0, k^1,\cdots,
                                                              \\
   (\quad)_{\le -1}:& \
       \text{projection onto } k^{-1}, k^{-2},\cdots.
                                                     \tag\eq  \\
\endalign
$$
These polynomials $\calB_n$ then satisfy the zero-curvature-like
equations
$$
  \dfrac{\rd \calB_m}{\rd t_n}
  - \dfrac{\rd \calB_n}{\rd t_m}
  + \{ \calB_m, \calB_n \} = 0.                      \tag\eq
$$
Thus one can reproduce the Lax formalism of the dispersionless KP
hierarchy.

In the case of our higher dimensional hierarchies, we have anther
canonical conjugate pair $(y,z)$ that commute with $(\hbar\rd_x,x)$,
hence altogether four operators $(\hbar\rd_x,x,y,z)$. These operators
obey the canonical commutation relations
$$
\align
  & [\hbar\rd_x,x] = \hbar, \quad [y,z] = \hbar,        \\
  & \text{other commutators} = 0.              \tag\eq  \\
\endalign
$$
In the limit of $\hbar \to 0$, these operators will be replaced by
canonical coordinates $(k,x,y,z)$ on a four dimensional symplectic
manifold, and by the standard rule
$$
   \hbar^{-1}[\quad,\quad] \longrightarrow \[ \quad, \quad \]
                                                        \tag\eq
$$
of quantum mechanics, the commutator of operators will be reduced
to the four dimensional Poisson bracket
$$
    \[ A, B \]   =   \frac{\rd A}{\rd k} \frac{\rd B}{\rd x}
                   - \frac{\rd A}{\rd x} \frac{\rd B}{\rd k}
                   + \frac{\rd A}{\rd y} \frac{\rd B}{\rd z}
                   - \frac{\rd A}{\rd z} \frac{\rd B}{\rd y}
                                                        \tag\eq
$$
of functions of $(k,x,y,z)$.  The role of Lax and Zakharov-Shabat
operators will thus be played by functions of $(k,x,y,z)$.
Those classical analogues of Lax and Zakharov-Shabat operators
will satisfy Poisson algebraic analogues of the Lax and
zero-curvature equations with respect to $\[\quad,\quad\]$.
Let us show details below.

\section{Hierarchy of commuting flows}

For the hierarchy of $t_{n\alpha}$ flows, classical counterparts
of $L$, $M$, $U$ and $V$ are given by the Laurent series
$$
\align
  \calL =& k + \sum_{n=1}^\infty g^{(0)}_{n+1} k^{-n},
                                                                    \\
  \calM =& \sum_{n,\alpha} n t_{n\alpha} \calL^{n-1} \calU^\alpha
        + x + \sum_{n=1}^\infty h^{(0)}_{n+1} \calL^{-n-1},
                                                                    \\
  \calU =& y + \sum_{n=1}^\infty u^{(0)}_n \calL^{-n},
                                                                    \\
  \calV =& \sum_{n,\alpha} \alpha t_{n\alpha} \calL^n \calU^{\alpha-1}
        + z + \sum_{n=1}^\infty v^{(0)}_n \calL^{-n},
                                                           \tag\eq  \\
\endalign
$$
where the coefficients $g^{(0)}_n$, $h^{(0)}_n$, $u^{(0)}_n$ and
$v^{(0)}_n$ are the leading terms of $\hbar$ expansion in
(\LimitOfGU) and (\LimitOfHV). These Laurent series satisfy
the canonical Poisson relations
$$
\align
  & \[ \calL, \calM \] = \[ \calU, \calV \] = 1, \\
  & \text{other Poisson brackets} = 0,     \tag\eq       \\
\endalign
$$
and the Poisson algebraic Lax equations
$$
\align
  \dfrac{\rd \calL}{\rd t_{n\alpha}}
                          = \[ \calB_{n\alpha}, \calL \], &\quad
  \dfrac{\rd \calM}{\rd t_{n\alpha}}
                          = \[ \calB_{n\alpha}, \calM \], \\
  \dfrac{\rd \calU}{\rd t_{n\alpha}}
                          = \[ \calB_{n\alpha}, \calU \], &\quad
  \dfrac{\rd \calV}{\rd t_{n\alpha}}
                          = \[ \calB_{n\alpha}, \calV \].
                                                  \tag\eq     \\
\endalign
$$
Here the classical counterparts $\calB_{n\alpha}$ of the
Zakharov-Shabat operators $B_{n\alpha}$ are now given by
$$
    \calB_{n\alpha} = \left( \calL^n \calU^\alpha \right)_{\ge 0},
                                                   \tag\eq
$$
where
$$
\align
    (\quad)_{\ge 0}:& \
        \text{projection onto } k^0, k^1, \cdots,                  \\
    (\quad)_{\le -1}:& \
        \text{projection onto } k^{-1}, k^{-2}, \cdots,
                                                          \tag\eq  \\
\endalign
$$
and satisfy the Poisson algebraic zero-curvature equations
$$
  \dfrac{\rd \calB_{m\alpha}}{\rd t_{n\beta}}
  - \dfrac{\rd \calB_{n\beta}}{\rd t_{m\alpha}}
  + \[ \calB_{m\alpha}, \calB_{n\beta} \] = 0.       \tag\eq
$$

Note that if we now impose the constraints
$$
\align
    & u^{(0)}_n = 0, \quad v^{(0)}_n = 0  \quad
      \text{for} \ n= 1,2,\ldots,                             \\
    \Bigl( \Longleftrightarrow
    & \quad
      \calU = y, \quad
      \calV = \sum \alpha t_{n\alpha} \calL^n y^{\alpha-1} + z
    \Bigr)
                                                    \tag\eq   \\
\endalign
$$
only the $t_{n0}$ flows remain nontrivial. The reduced hierarchy
is substantially the same as the dispersionless KP hierarchy.

\section{Hierarchy of noncommuting flows}

For the hierarchy of $(t,p,q)$ flows, classical counterparts of
$L$, $U$ and $V$ are given by
$$
\align
  \calL =& k + \sum_{n=1}^\infty g^{(0)}_{n+1}k^{-n},          \\
  \calU =& \sum_{n=1}^\infty p_n \calL^n + y
         + \sum_{n=1}^\infty u^{(0)}_n \calL^{-n},             \\
  \calV =& \sum_{n=1}^\infty q_n \calL^n + z
         + \sum_{n=1}^\infty v^{(0)}_n \calL^{-n}.    \tag\eq  \\
\endalign
$$
They satisfy the Poisson bracket relations
$$
   \[ \calL, \calU \] = \[ \calL, \calV \] = 0, \quad
   \[ \calU, \calV \] = 1,                        \tag\eq
$$
and the Poisson algebraic Lax equations
$$
\align
  & \hbar \dfrac{\rd \calL}{\rd t_n} = \[\calB_n, \calL\],  \quad
    \hbar \dfrac{\rd \calL}{\rd p_n} = \[\calC_n, \calL\],  \quad
    \hbar \dfrac{\rd \calL}{\rd q_n} = \[\calD_n, \calL\],       \\
  & \hbar \dfrac{\rd \calU}{\rd t_n} = \[\calB_n, \calU\],  \quad
    \hbar \dfrac{\rd \calU}{\rd p_n} = \[\calC_n, \calU\],  \quad
    \hbar \dfrac{\rd \calU}{\rd q_n} = \[\calD_n, \calU\],       \\
  & \hbar \dfrac{\rd \calV}{\rd t_n} = \[\calB_n, \calV\],  \quad
    \hbar \dfrac{\rd \calV}{\rd p_n} = \[\calC_n, \calV\],  \quad
    \hbar \dfrac{\rd \calV}{\rd q_n} = \[\calD_n, \calV\].
                                                      \tag\eq \\
\endalign
$$
where $\calB_n$, $\calC_n$ and $\calD$ are given by
$$
\align
  \calB_n =& \bigl( \calL^n \bigr)_{\ge 0},                         \\
  \calC_n =& \Bigl( - \calV \calL^n + \frac{1}{2}\sum_{m=1}^\infty
                           q_m \calL^{m+n} \Bigr)_{\ge 0},          \\
  \calD_n =& \Bigl(   \calU \calL^n - \frac{1}{2}\sum_{m=1}^\infty
                           p_m \calL^{m+n} \Bigr)_{\ge 0},
                                                            \tag\eq \\
\endalign
$$
and satisfy the Poisson algebraic zero-curvature equations
$$
\align
    \dfrac{\rd \calB_m}{\rd t_n}
    -  \dfrac{\rd \calB_n}{\rd t_m}
    + \[ \calB_m, \calB_n \] =& 0,
                                                              \\
    \dfrac{\rd \calB_m}{\rd p_n}
    -  \dfrac{\rd \calC_n}{\rd t_m}
    + \[ \calB_m, \calC_n \] =& 0,
                                                              \\
    \dfrac{\rd \calB_m}{\rd q_n}
    -  \dfrac{\rd \calD_n}{\rd t_m}
    + \[ \calB_m, \calD_n \] =& 0,
                                                              \\
    \dfrac{\rd \calC_m}{\rd p_n}
    -  \dfrac{\rd \calC_n}{\rd p_m}
    + \[ \calC_m, \calC_n \] =& 0,
                                                              \\
    \dfrac{\rd \calC_m}{\rd q_n}
    -  \dfrac{\rd \calD_n}{\rd p_m}
    + \[ \calC_m, \calD_n \] =& 0,
                                                              \\
    \dfrac{\rd \calD_m}{\rd q_n}
    -  \dfrac{\rd \calD_n}{\rd q_m}
    + \[ \calD_m, \calD_n \] =& 0.
                                                    \tag\eq   \\
\endalign
$$

Selfdual gravity can now be reproduced under the constraints
$$
    g^{(0)}_{n+1} = 1 \quad \text{for} \ n = 1,2,\ldots
    \quad \Bigl( \Longleftrightarrow \calL = k \Bigr).
                                                          \tag\eq
$$
All $t_n$ flows then become trivial, and upon identifying
$\lambda = k$, the reduced hierarchy of remaining $(p,q)$ flows
reproduces the same hierarchy as constructed in Ref. \TaSDG\
for the second heavenly equation. Note that in the previous section,
we first imposed constraints, then took the $\hbar \to 0$ limit.
In this section, we first took the $\hbar \to 0$ limit, then put
the constraints.  We have thus two routes to the same
second heavenly equation.

The status of the first heavenly equation is rather obscure from
this point view. As mentioned in the last section, a better approach
will be to start from a Toda version of the hierarchy of $(t,p,q)$
flows, and to consider quasi-classical limit. Similarly, a Toda
version of the hierarchy of $t_{n\alpha}$ flows will yield,
in the same quasi-classical limit, a higher dimensional extension
of the dispersionless Toda hierarchy
[\REF\TTdToda{
  Takasaki, K., and Takebe, T.,
  SDiff(2) Toda equation ---
  hierarchy, tau function and symmetries,
  Lett. Math. Phys. 23 (1991), 205-214.
}\TTdToda].

\section{Moyal vs. Poisson algebraic hierarchies}

Remarkably, unlike the Moyal algebraic hierarchies, these Poisson
algebraic hierarchies are systems of ``algebraic differential
equations" --- they consist of an infinite number of equations,
but each equation contains only a finite number of terms.  This
is not the case for the Moyal algebraic hierarchy, because the
Moyal bracket itself is an infinite power series of $\hbar$ and
contains arbitrarily high order derivatives of the unknown
functions. First of all, Moyal algebraic deformations of the
two heavenly equations themselves are non-algebraic in this sense.

A drawback is that the dressing operator method does not work
directly within the Poisson algebraic hierarchies. The dressing
operator method can be used only via the Moyal algebraic
hierarchies. Twistor theory, which has been successful for
selfdual gravity and the dispersionless KP and Toda hierarchies
[\TaSDG][\TTdKP][\TTdToda], will provide an alternative approach.

\chapter{Conclusion --- overview from W-infinity algebras}

\noindent
Inspired by the philosophy of large-$N$ limit, we have constructed
higher dimensional analogues of the KP hierarchy.  Conceptually,
the new hierarchies may be interpreted as large-$N$ limit of the
$N$-component KP hierarchy.  We, however, have avoided to deal
with finite-$N$ models and given a direct construction of such
hierarchies. Fundamental constituents of the hierarchies are
pseudo-differential operators with Moyal algebraic coefficients.

Actually, we have found two different types hierarchies consisting,
respectively, of commuting flows and noncommuting flows. The hierarchy
of commuting flows resembles the ordinary and multi-component KP
hierarchies, whereas the hierarchy of noncommuting flows includes
Moyal algebraic deformations of selfdual gravity as a reduction.

Furthermore, both the commuting and noncommuting hierarchies have
turned out to possess quasi-classical limit, whose Lax formalism
is based on Poisson brackets rather than Moyal brackets. In the
quasi-classical limit, the commuting hierarchy gives a higher
dimensional extension of the dispersionless KP hierarchy.  The
noncommuting hierarchy, meanwhile, includes ordinary selfdual
gravity as a reduction.

It is instructive to review our results from the point of view
of W-infinity algebras
[\REF\WinfGen{
  Bakas, I.,
  The structure of the $W_\infty$ algebra,
  Commun. Math. Phys. 134 (1990), 487-508. \nextline
  Pope, C.N., Romans, L.J., and Shen, X.,
  The complete structure of $W_\infty$,
  Phys. Lett. 236B (1990), 173-178. \nextline
  Pope, C.N., Romans, L.J., and Shen, X.,
  Ideals of Kac-Moody algebras and realizations of $W_\infty$,
  Phys. Lett. 245B (1990), 72-78.
}\WinfGen].
The following is a list of integrable systems that fall into our
current scope.
\medskip
\item{(0)} ordinary KP hierarchy (KP),
\item{(1)} $N$-component KP hierarchy ($N$-KP),
\item{(2)} nonabelian KP hierarchies with Moyal algebraic
coefficients (MAKP),
\item{(3)} Moyal algebraic deformations of selfdual gravity (MASDG),
\item{(4)} $N$-component AKNS-ZS hierarchy ($N$-AKNSZS).
\medskip
\noindent
These integrable systems are linked with each other as the following
diagram indicates.
$$
\matrix
  \text{KP} & \mapright{N-\text{comp.}} & N-\text{KP}
            & \mapright{N \to \infty}   & \text{MAKP}              \\
            &                    & \mapdown{\hbar\rd_x \to \lambda}
            &                    & \mapdown{\hbar\rd_x \to \lambda}\\
            &                           & \text{$N$-AKNSZS}
            & \mapright{N \to \infty}   & \text{MASDG}             \\
\endmatrix
$$
[Remark: More precisely, large-$N$ limit of $N$-AKNSZS gives
a reduction of the commuting MAKP hierarchy rather than of the
noncommuting MAKP hierarchy, hence the last line of the above
diagram is somewhat inaccurate.] The following list shows
infinite dimensional Lie algebras associated with these
integrable systems.
\medskip
\item{(0)}
$W_\infty = \left< x^i (\hbar\rd_x)^j \right>$ \
(ordinary $W_\infty$ algebra [\WinfGen]),

\item{(1)}
$W^N_\infty = \left< E_{\alpha\beta} x^i (\hbar\rd_x)^j \right>$ \
($W_\infty$ algebra with inner symmetries
[\REF\WNinf{
  Bakas, I., and Kiritsis, E.,
  Grassmannian coset models and unitary representation of $W_\infty$,
  Mod. Phys. Lett. A5 (1990), 2039-2050.\nextline
  Odake, S., and Sano, T.,
  $W_{1+\infty}$ and super $W_\infty$ algebras with SU($N$) symmetry,
  Phys. Lett. B258 (1991), 369-374.
}\WNinf]),

\item{(2)}
$W^\infty_\infty = \left< y^\alpha z^\beta x^i (\hbar\rd_x)^j \right>$ \
(large-$N$ limit of $W^N_\infty$ [\WNinf]),

\item{(3)}
$\calL \Moyal(\Sigma) = \left< y^\alpha z^\beta \lambda^j \right>$ \
(loop algebra of Moyal algebra),

\item{(4)} $\calL \gl(N) = \left< E^{\alpha\beta} \lambda^j \right>$ \
(loop algebra of $\gl(N)$),
\medskip
\noindent
where $<\cdots>$ shows generators; $E^{\alpha\beta}$ denote the
standard generators of $\gl(N)$,
$$
    \bigl( E^{\alpha\beta} \bigr)_{ij}
    = \delta_{i\alpha} \delta_{j\beta}.                       \tag\eq
$$
The first four of these algebras are of W-infinity type, whereas
the last one is of Kac-Moody type.  The above diagram of integrable
systems simultaneously show the interrelation of these Lie algebras:
$$
\matrix
  W_\infty  & \mapright{N-\text{comp.}} & W^N_\infty
            & \mapright{N \to \infty}   & W^\infty_\infty          \\
            &                    & \mapdown{\hbar\rd_x \to \lambda}
            &                    & \mapdown{\hbar\rd_x \to \lambda}\\
            &                           & \calL\gl(N)
            & \mapright{N \to \infty}   & \calL\Moyal(\Sigma)      \\
\endmatrix
$$
In the limit of $\hbar \to 0$, these W-infinity algebras are
contracted to classical counterparts.  In particular,
$\calL\Moyal(\Sigma)$ and $W^\infty_\infty$ turn into
Poisson algebras $\calL\Poisson(\Sigma)$
and $w^\infty_\infty \simeq \Poisson(M)$, $\dim M = 4$.
This is exactly what we have seen in the transition from
Moyal algebraic hierarchies to Poisson algebraic hierarchies.

We conclude this paper with a list of issues to be pursued
along the same line of approach. These issues will be discussed
elsewhere.

(a) {\it Toda version of MAKP}.
We noted in a previous section that a Toda version of MAKP
will give a natural framework for dealing with the first
heavenly equation of selfdual gravity. This should be a
rather straightforward extension, simply replacing
pseudo-differential operators by difference operators
(with Moyal algebraic coefficients).

(b) {\it More detailed study of Poisson algebraic
hierarchies}.
The Poisson algebraic hierarchies will provide new material
for the twistor theoretical approach to nonlinear integrable
systems. Furthermore, as the dispersionless KP hierarchy
is applied to Landau-Ginzburg models of topological strings
[\REF\DVV{
  Dijkgraaf, R., Verlinde, E., and Verlinde, H.,
  Topological strings in $d < 1$,
  Nucl. Phys. B352 (1991), 59-86.
}\DVV],
the higher dimensional hierarchies might be related to
similar physical models.

(c) {\it Multi-component version of MAKP}.
This extension is also straightforward, just redefining
the pseudo-differential operators $W$, $L$, etc. to have
$N \times N$ matrix valued coefficients with Moyal algebraic
matrix elements.  According to a preliminary analysis
(we omit details), a hierarchy of Bogomolny-type
[\REF\BogoGen{
  Mason, L.J., and Sparling, G.A.J.,
  Twistor correspondences for the soliton hierarchies,
  J. Geom. Phys. 8 (1992) 243-271.
}\BogoGen]
can be derived from this $N$-component version ($N$-MAKP) as
a reduction. Thus the previous diagram can further be enlarged
as:
$$
\matrix
  \text{MAKP} & \mapright{N-\text{comp.}} & N-\text{MAKP}    \\
              &                 & \mapdown{\text{reduction}} \\
              &                           & \text{Bogomolny} \\
\endmatrix
$$
This reduction is however somewhat strange --- constraints
are imposed on both the dressing operator {\it and} the time
variables.  This interpretation of Bogomolny hierarchies
resembles the ``$D$-module approach" of Sato et al.
[\REF\Dmodule{
  Sato, M.,
  The KP hierarchy and infinite dimensional Grassmann manifolds,
  in: Proc. Symp. Pure Math. 49
  (American Mathematical Society, 1989). \nextline
  Takasaki, K.,
  Integrable systems as deformations of ${\cal D}$-modules,
  ibid.\nextline
  Ohyama, Y.,
  Self-duality and integrable systems,
  thesis (Kyoto University, 1990).
}\Dmodule].
Furthermore, the fact that the time variables, too, have to be
constrained is somewhat reminiscent of the ``small phase space"
of Landau-Ginzburg models [\DVV].

(d) {\it Hierarchies with compactified Moyal algebras}.
In the case of a torus, the Moyal algebra (or ``quantum torus
algebra") has an invariant trace
[\REF\quantumTorus{
  Hoppe, J., Olshanetsky, M., and Theisen, S.,
  Dynamical systems on quantum tori algebras,
  Karlsruhe preprint KA-THEP-10/91 (October, 1991).
}\quantumTorus]
This will allow us to construct a higher dimensional tau function,
because the usual trace of $\gl(N)$ plays an implicit but substantial
role in the theory of tau functions of multi-component KP hierarchies.

The author is very grateful to Jens Hoppe, Stanislav Pakuliak,
Ian Strachan and Takashi Takebe for a number of useful comments.
This work is partly supported by the Grant-in-Aid for Scientific
Research, the Ministry of Education, Science and Culture, Japan.

\refout
\bye